\documentclass[a4paper,12pt]{article}
\pdfoutput=1

\usepackage{graphicx,hyperref,xcolor}
\usepackage{amsmath,amssymb}
\usepackage{latexsym}
\usepackage{cite}

\usepackage[normalem]{ulem}
\usepackage{cancel}
\usepackage{setspace}

\hypersetup{
bookmarksnumbered=true,
pdfborder={0 0 0},
bookmarkstype=toc,
colorlinks={true},
linkcolor={blue},
urlcolor={blue},
filecolor={blue},
citecolor={blue}
}

\definecolor{refkey}{rgb}{0.40, 0.55, 0.55}
\definecolor{labelkey}{rgb}{0.40, 0.55, 0.55}

\usepackage[a4paper,top=30truemm,left=20truemm,textwidth=170truemm,textheight=237truemm]{geometry}

\makeatletter
  
  \@addtoreset{equation}{section}
\makeatother

\usepackage{comment}

\newcommand{\TildeBox}{\raisebox{2pt}{$\widetilde{\raisebox{-2pt}{$\Box$}}$}}

\begin{document}
\allowdisplaybreaks{
\begin{titlepage}
\null
\begin{flushright}
March, 2018 \\
TIT/HEP-664
\end{flushright}

\vskip 1.8cm
\begin{center}

  {\Large \bf Worldsheet Instanton Corrections to}
\vskip 0.3cm
 {\Large \bf Five-branes and Waves in Double Field Theory
}

\vskip 1.8cm
\normalsize

  {\bf Tetsuji
 Kimura\footnote{tetsuji(at)th.phys.titech.ac.jp}{}$^{\text{,}}$\footnote{The
 affiliation since April 2018: Research Institute of Science and
 Technology, College of Science and Technology, Nihon University, 1-8-14
 Kanda Surugadai, Chiyoda-ku, Tokyo 101-8308, Japan.},
Shin Sasaki\footnote{shin-s(at)kitasato-u.ac.jp}
and Kenta Shiozawa\footnote{k.shiozawa(at)sci.kitasato-u.ac.jp}}

\vskip 0.5cm

  { \it
  ${}^1$Department of Physics \\
  Tokyo Institute of Technology \\
  Tokyo 152-8551, Japan \\
\vspace{0.5cm}
  ${}^{3,4}$Department of Physics \\
  Kitasato University \\
  Sagamihara 252-0373, Japan \\
  }

\vskip 1.5cm

\begin{abstract}
We make a comprehensive study on the string winding corrections to supergravity
 solutions in double field theory (DFT).
We find five-brane and wave solutions of diverse codimensions in which the winding coordinates are naturally included.
We discuss a physical interpretation of the winding coordinate dependence.
The analysis based on the geometric structures behind the solutions
 leads to an interpretation of the winding dependence as string worldsheet instanton corrections.
We also give a brief discussion on the origins of these winding corrections in gauged linear sigma model.
Our analysis reveals that for every supergravity solution, one has DFT
 solutions that include string winding corrections.
\end{abstract}
\end{center}
\end{titlepage}

\newpage
\tableofcontents
\setcounter{footnote}{0}

\section{Introduction}
Understanding the space-time structure in the Planck scale is one of the most
important theme in theoretical physics.
One knows that a macroscopic nature of the universe is captured by general
relativity where the space-time is probed by a point particle.
The geometry of space-time is realized as a Riemannian manifold whose
metric satisfies the Einstein equation.
One would expect that this picture fails in the Planck scale where quantum mechanical effects
drastically change the structure of space-time.
String theory, which is a candidate of consistent quantum gravity,
provides rather intrinsic nature of the Planck scale space-time.
Since space-time is probed by a one-dimensional extended object ({\it
i.e.} the fundamental string (F-string)) it is therefore plausible that
the conventional Riemannian geometry is modified in string theory.

One of the most important
features 
of string theory is U-duality.
Consider M-theory compactified on $d$-dimensional torus $T^d$, the
low-energy supergravity theories in $D=11-d$ space-time
dimensions have U-duality hidden symmetry based on the $E_{d(d)}$ group \cite{Obers:1998fb}.
Among other things, the existence of T-duality which interchanges the Kaluza-Klein (KK) and
winding modes of wrapped strings is the most prominent difference
between the theories based on point particles and strings.
This makes it expressive that the T-duality interchanges the
``geometrical'' (or conventional) coordinates
$x^\mu$, associated with the Fourier dual of the KK-modes, and the
``winding'' coordinates $\tilde{x}_\mu$ associated with the winding modes.

According to the T-duality symmetry in a setup of the flux
compactification, a notion of non-geometric fluxes is revealed
\cite{Hellerman:2002ax}.
This is rephrased as a T-duality chain of the $H$-$\tau$-$Q$-$R$ fluxes
\cite{Shelton:2005cf}.
On the other hand, there is yet another T-duality chain of five-branes in type
II string theories.
The H-monopole, which is obtained by compactifying a transverse direction to
the NS5-brane to $S^1$, has a $U(1)$ isometry. The T-duality
transformation of the H-monopole along the isometry direction results in the KK-monopole
(KK5-brane) which is geometrically described by the Taub-NUT space.
As a consequence, the $H$-flux sourced by the H-monopole is
mapped to the geometric $\tau$-flux sourced by the KK-monopole.
It is notable that we can perform further T-duality transformations.
By introducing an extra isometry along a transverse direction to the
KK-monopole, the second T-duality transformation becomes possible.
The resulting solution is known as the $Q$-brane.
This is a kind of exotic branes appearing non-trivially in the T-duality orbit.
The conventional argument labels this as the $5^2_2$-brane \cite{Obers:1998fb}.
Most notably, the $5^2_2$-brane is a non-geometric object in the sense that
its background geometry has a monodromy in the T-duality group $O(2,2)$ \cite{deBoer:2010ud}. This kind of non-geometric backgrounds is called
the T-fold \cite{Hull:2006va}.
The $Q$-brane is a source of the $Q$-flux.
Indeed, exotic branes are sources of non-geometric fluxes \cite{Hassler:2013wsa,Sakatani:2014hba,Andriot:2014uda}.
A further T-duality transformation settles down to the yet unknown $R$-brane
($5^3_2$-brane) which is the source of the $R$-flux.

These famous T-duality chains cause an interesting observation.
For example, let us focus on the T-duality relation between the H- and
KK-monopoles.
In \cite{Tong:2002rq}, string worldsheet instanton corrections to
 the $S^1$-smeared NS5-brane ({\it i.e.} the H-monopole) geometry
is studied through the gauged linear sigma model (GLSM).
As a result, the instanton corrections break the isometry in the $S^1$
and the H-monopole becomes the NS5-brane localized in the $S^1$.
A puzzle arises in the T-dualized KK-monopole side.
Since the Taub-NUT geometry inherits the isometry, what is the corresponding worldsheet
instanton effect?
The worldsheet instanton corrections to the KK-monopole
is analyzed \cite{Harvey:2005ab, Jensen:2011jna} and
it is shown that the corrections break the isometry in the {\it winding
(T-dualized) space}. In other words, the worldsheet instanton
corrections in the KK-monopole geometry introduces the dual
``winding'' coordinates $\tilde{x}_\mu$ dependence in the solution in addition to the space-time
geometrical coordinates $x^\mu$.
This originates essentially from an old question for non-isometric T-duality in the unwinding
string process \cite{Gregory:1997te}.
The same is true in further T-duality processes.
Based on the GLSM introduced in \cite{Kimura:2013fda},
two of the present authors studied the worldsheet instanton corrections
to the $5^2_2$-brane geometry \cite{Kimura:2013zva} where again the
winding coordinate dependence appears via the instanton effects.

Double field theory (DFT) \cite{Hull:2009mi} is a formalism where the T-duality group
$O(d,d)$ is realized as a manifest symmetry\footnote{
A manifestly T-duality covariant formalism is initially developed in \cite{Siegel:1993xq}.}.
The T-duality transformation is encoded into the generalized coordinate
transformations in doubled space $M^{2D}$ with the doubled coordinates $X^M = (x^\mu, \bar{x}_\mu)$.
Recent studies reveal that the DFT contains interesting classical solutions.
Among other things, the supergravity solutions of the F-string and
its T-dualized wave are shown to be realized as the DFT wave solution
\cite{Berkeley:2014nza}.
Soon after the discovery of the DFT wave solution, it is found that the H-monopole and its
T-dualized KK-monopole are embedded in the DFT monopole solution \cite{Berman:2014jsa}.
A subsequent work finds that the non-geometric $Q$- and $R$-branes are also
encoded in the DFT monopole solution \cite{Bakhmatov:2016kfn}.

It is worthwhile to emphasize that one can perform the T-duality
transformations in DFT without assuming isometries.
Therefore, the generalized T-duality without isometry first mentioned in
\cite{Dabholkar:2005ve} is naturally realized in the DFT framework.
This fact means that locally non-geometric structures based on the dual
coordinate dependence of the geometry are allowed as consistent string
backgrounds in DFT.
Indeed, it is shown that the localized KK-monopole in the winding space is
a solution to DFT \cite{Berman:2014jsa}.
The winding dependence of the solution is interpreted as non-geometric
corrections to the ordinary KK-monopole solution in supergravity.
The corrections appeared in the DFT solution precisely agree with the worldsheet
 instanton corrections studied in \cite{Harvey:2005ab}.
The same discussion holds for the localized $5^2_2$-brane in the winding
space \cite{Bakhmatov:2016kfn} which is again consistent with the result
found in \cite{Kimura:2013zva}.
The winding coordinate dependence of the KK-monopole and
 $5^2_2$-brane solutions is mentioned in \cite{VallCamell:2017ogl}.
These facts suggest that the worldsheet instanton corrections to
space-time geometries are naturally included in the classical DFT solutions.

In this paper, we give a systematic analysis on
the winding corrections to the five-brane geometries via DFT solutions.
We explicitly write down all the solutions of five-branes of codimensions
$d = 4,3,2,1$ appearing in the T-duality chains and study the string winding corrections to them.
We provide qualitative discussions on the interpretation of the winding corrections as string worldsheet instanton effects.
As a byproduct we discuss string winding corrections to the wave and F-string solutions.
We also perform the last T-duality transformation and discuss
the space-filling five-brane in six dimensions that is fully localized
in the winding space.

The organization of this paper is as follows.
In the next section, we introduce the DFT action and write down the
equations of motion.
We then solve the DFT equations by using the five-brane and
wave ans\"{a}tze.
We explicitly show that there are variety of solutions which depend on
winding coordinates.
In section \ref{sect:resum}, we rewrite the solutions to the forms which
enable one to interpret the winding
coordinate dependence as corrections to the
supergravity solutions.
In section \ref{sect:instanton}, we provide a possible interpretation of
the solutions as worldsheet instanton corrections.
In section \ref{sect:GLSM} we discuss GLSM interpretation of the winding corrections.
Section \ref{sect:conclusion} is devoted to conclusion and discussions.

\section{DFT solutions of diverse codimensions}\label{sect:DFT_solutions}
In this section, we discuss five-brane solutions in DFT.
The fundamental fields in DFT are the generalized metric
$\mathcal{H}_{MN}$ and the generalized dilaton $d$ defined by \cite{Hull:2009mi}:
\begin{align}
\mathcal{H}_{MN} =
\left(
\begin{array}{cc}
g_{\mu \nu} - B_{\mu \rho} g^{\rho \sigma}  B_{\sigma \nu} & B_{\mu \rho} g^{\rho \nu} \\
- g^{\mu \rho} B_{\rho \nu} & g^{\mu \nu}
\end{array}
\right)
,
\qquad
e^{-2d} = \sqrt{-g} \, e^{-2\phi}.
\label{eq:generalized_metric}
\end{align}
Here all the fields in DFT are defined in the doubled space
$X^M = (x^\mu,\bar{x}_\mu) \ (\mu = 1, \ldots, d)$.
The components $g_{\mu \nu}$, $B_{\mu \nu}$ and $\phi$ are reduced to the metric and
the Kalb-Ramond $B$-field and the dilaton in a certain supergravity frame
after the strong constraint is imposed\footnote{
The strong constraint in DFT is defined by
$\eta^{MN} \partial_M * \partial_N * = $ for any fields and gauge
parameters $*$. Here $\partial_M = \frac{\partial}{\partial X^M}$, and
$\eta^{MN}$ is defined in \eqref{eq:2dinvmetric}.
}.
The generalized metric is parametrized by the
$O(d,d)/(O(d) \times O(d))$ coset space
and whose inverse $\mathcal{H}^{MN}$ is defined by
\begin{align}
\mathcal{H}^{MP} \mathcal{H}_{PN} = \delta^M {}_N.
\end{align}
The
inverse of the generalized metric is obtained
through the uplifting of the $O(d,d)$ indices:
\begin{align}
\mathcal{H}^{MN} = \eta^{MP} \eta^{NQ} \mathcal{H}_{PQ}.
\end{align}
This is a consequence of the fact that $\mathcal{H}_{MN}$ is an element
of $O(d,d)$.
Here $\eta_{MN}$ and $\eta^{MN}$ are the $2d \times 2d$, $O(d,d)$ invariant metric and its inverse
defined by
\begin{align}
\eta_{MN} =
\left(
\begin{array}{cc}
0 & \delta_{\mu} {}^{\nu} \\
\delta^{\mu} {}_{\nu} & 0
\end{array}
\right)
,
\qquad
\eta^{MN} =
\left(
\begin{array}{cc}
0 & \delta^{\mu} {}_{\nu} \\
\delta_{\mu} {}^{\nu} & 0
\end{array}
\right)
.
\label{eq:2dinvmetric}
\end{align}
In the following, we always raise and lower the $O(d,d)$ indices by
the $O(d,d)$ invariant metrics $\eta_{MN}, \eta^{MN}$.
The DFT action is given by \cite{Hohm:2010pp}:
\begin{align}
S_{\text{DFT}} = \int \! d^{2d} X \, e^{-2d} \, \mathcal{R} (\mathcal{H}, d).
\label{eq:DFT_action}
\end{align}
Here the quantity $\mathcal{R}$ is constructed from $\mathcal{H}$ and
$d$ and is invariant under the generalized coordinate transformation.
This is called the generalized Ricci scalar and given by
\begin{align}
\mathcal{R} =& \
4 \mathcal{H}^{MN} \partial_M \partial_N d - \partial_M \partial_N
 \mathcal{H}^{MN} - 4 \mathcal{H}^{MN} \partial_M d \partial_N d
+ 4 \partial_M \mathcal{H}^{MN} \partial_N d
\notag \\
&
+ \frac{1}{8}
 \mathcal{H}^{MN} \partial_M \mathcal{H}^{KL} \partial_N
 \mathcal{H}_{KL}
 - \frac{1}{2} \mathcal{H}^{MN} \partial_M
 \mathcal{H}^{KL} \partial_K \mathcal{H}_{NL}.
\label{eq:generalized_Ricci_scalar}
\end{align}
If we move to the supergravity frame by solving the strong constraint with the condition
$\bar{\partial}_{\mu} = 0$ the DFT action
is reduced to the bosonic part of the supergravity action in the NSNS sector
\begin{align}
S = \int \! d^D x \, \sqrt{-g} \, e^{-2\phi}
\left[
R + 4 (\partial \phi)^2 - \frac{1}{12} (H^{(3)})^2
\right],
\label{eq:NSNS_gravity}
\end{align}
where $R$ is the Ricci scalar defined by the metric $g_{\mu \nu}$
and $H^{(3)} = d B$ is the field strength of the Kalb-Ramond field.

The variation of the action
\eqref{eq:DFT_action} with respect to the generalized metric and
the generalized dilaton results in the equations of motion
\begin{align}
\delta d  : \ \mathcal{R} = 0, \qquad
\delta \mathcal{H}^{MN}  : \ P_{MN} {}^{KL} {\mathcal K}_{KL} = 0.
\label{eq:eom}
\end{align}
Here $P$ is a projection given by
\begin{align}
P_{MN} {}^{KL} =
\frac{1}{2} \left[
\delta_M {}^{(K} \delta_N {}^{L)} - \mathcal{H}_{MP} \eta^{P (K}
 \eta_{NQ} \mathcal{H}^{L) Q}
\right],
\label{eq:projection}
\end{align}
where the symmetrization is defined by
$A_{(M} B_{N)} = (A_M B_N + A_N B_M)/2$.
The $O(d,d)$ tensor $\mathcal{K}_{MN}$ is defined by \cite{Hohm:2010pp}:
\begin{align}
{\mathcal K}_{MN} &=
	{1 \over 8} \partial_M {\mathcal H}^{KL} \partial_N {\mathcal H}_{KL}
	- {1 \over 4} (\partial_L - 2\partial_L d) {\mathcal H}^{KL} \partial_K {\mathcal H}_{MN}
	+ 2 \partial_M \partial_N d
\notag \\
& \quad
	- {1 \over 2} \partial_{(M} {\mathcal H}^{KL} \partial_L {\mathcal H}_{N)K}
	+ {1 \over 2} (\partial_L - 2\partial_L d) \left[ {\mathcal H}^{KL} \partial_{(M} {\mathcal H}_{N)K}
		+ \eta^{KP} \eta^{LQ} {\mathcal H}_{P(M} \partial_K {\mathcal H}_{N)Q} \right].
\end{align}
We call this the $\mathcal{K}$-tensor.
In the following, we explore five-brane and wave solutions to these equations.

\subsection{Five-brane solutions}
We first focus on the five-brane solutions.
It is convenient to classify brane solutions by codimensions.
In the following, we examine five-brane solutions of codimensions
4,3,2,1 and a space-filling brane.

\subsubsection{Codimension four}
We first start from the five-branes of codimension four.
The generalized coordinates are parametrized by the longitudinal
coordinates
$x^m \ (m=0,\ldots, 5)$ and the transverse coordinates $y^a \ (a = 6,\ldots,9)$ of the
five-branes. Together with their doubled pairs
$(\bar{x}_m, \bar{y}_a)$\footnote{
Note that the coordinates
$\bar{x}_m$, $\bar{y}_a$ simply stand for the doubled
pairs but not the winding coordinates. In the following, we will assign
the roles of geometrical and winding coordinates
to each coordinate $(x^m, y^a, \bar{x}_m, \bar{y}_a)$.
},
they are organized into the doubled coordinates:
\begin{align}
X^M =
(x^m, y^a; \bar{x}_m, \bar{y}_a).
\end{align}
We employ the ansatz for the localized DFT monopole solution
\cite{Berman:2014jsa}:
\begin{align}
ds^2_{\text{DFT}}
&= H (\delta_{ab} - H^{-2} b_{ac} b^c{}_b) dy^a dy^b
+ H^{-1} \delta^{ab} d\bar{y}_a d\bar{y}_b
+ 2 H^{-1} b_a{}^b dy^a d\bar{y}_b
\notag \\
&\quad
+ \eta_{mn} dx^m dx^n + \eta^{mn} d\bar{x}_m d\bar{x}_n,
\notag \\
d
&
= \text{const.} - \frac{1}{2} \log H,
\label{eq:DFT_monopole_ansatz}
\end{align}
where the DFT ``line element'' is understood as
$ds^2_{\text{DFT}} = \mathcal{H}_{MN} dX^M dX^N$.
$H$ and $b_{ab}$ will satisfy certain conditions.
More explicitly, the components of the
generalized metric are
\begin{align}
{\mathcal H}_{ab} &= H(\delta_{ab} - H^{-2} b_{ac} b^c{}_b),
& {\mathcal H}_{\bar{a}\bar{b}} &= H^{-1} \delta_{\bar{a}\bar{b}},
&{\mathcal H}_{a\bar{b}} &= H^{-1} b_{a\bar{b}},
\notag \\
{\mathcal H}_{mn} &= \eta_{mn},
& {\mathcal H}_{\bar{m}\bar{n}} &= \eta_{\bar{m}\bar{n}},
& {\mathcal H}_{\bar{a}b} &= - H^{-1} b_{\bar{a}b}, \\
\notag \\
{\mathcal H}^{ab} &= H^{-1} \delta^{ab},
& {\mathcal H}^{\bar{a}\bar{b}} &= H(\delta^{\bar{a}\bar{b}}
	- H^{-2} b^{\bar{a}c} b_c{}^{\bar{b}}),
& {\mathcal H}^{a\bar{b}} &= - H^{-1} b^{a\bar{b}},
\notag \\
{\mathcal H}^{mn} &= \eta^{mn},
& {\mathcal H}^{\bar{m}\bar{n}} &= \eta^{\bar{m}\bar{n}},
& {\mathcal H}^{\bar{a}b} &= H^{-1} b^{\bar{a}b},
\label{eq:DFT_monopole}
\end{align}
where $a = 6,7,8,9$.
The index with bar is an alternative notation for the
dual coordinates $(\bar{x}^{\bar{m}}, \bar{y}^{\bar{a}})$.
The solution is completely determined by the function $H$.
For codimension four solutions, we assume that
the function depends on the four-dimensional transverse
directions $H = H(y^a)$.
Since the solution does not depend on both the doubled pairs
of coordinates simultaneously, it satisfies the strong constraint.
It is important to bear in mind that we never say that $y^a$ are
geometrical coordinates. It depends on the frame which we consider as we will discuss below.

Now we examine the conditions for the function $H$.
The first equation of motion in \eqref{eq:eom} is the vanishing
condition of the generalized Ricci scalar $\mathcal{R}$.
Substituting the ansatz \eqref{eq:DFT_monopole_ansatz}
into the definition of $\mathcal{R}$ \eqref{eq:generalized_Ricci_scalar}, we obtain
\begin{align}
\mathcal{R} =
 \frac{1}{2} H^{-3} \delta^{ab} \, \partial_a H \partial_b H
- \frac{1}{4} H^{-3} \delta^{ab} \, \partial_a b^{cd} \, \partial_b b_{cd}
+ \frac{1}{2} H^{-3} \partial_a b^{cd} \, \partial_c b^a{}_d
- H^{-2} \delta^{ab} \, \partial_a \partial_b H.
\end{align}
In order for this to vanish, we impose the condition
\begin{align}
3 \partial_{[a} b_{bc]}
= \varepsilon_{abcd} \, \partial^d H(y),
\qquad
\label{eq:BPS_condition}
\end{align}
where $\varepsilon_{abcd}$ is the totally anti-symmetric symbol.
This condition results in the equation $\Box H = 0$: $\Box$ is the
four-dimensional Laplacian defined in the space along $y^a$.
It turns out that $H$ is the harmonic function.

Next, we consider the second equation in \eqref{eq:eom}.
For the ansatz \eqref{eq:DFT_monopole_ansatz}, the non-zero components of the
$\mathcal{K}$-tensor are calculated to be
\begin{align}
{\mathcal K}_{ab}
&= {1 \over 2} H^{-4} \Big[ \delta^{cd} \, b_{ac} b_{db} \, \delta^{ef} \partial_e H \partial_f H
		+ 2 b_a{}^c b_b{}^d \, \partial_c H \partial_d H \Big]
\notag \\
& \quad
	- {1 \over 4} H^{-3} \Big[
		b_{bc} \, \partial_a H \, \partial_d b^{cd}
		+ b_{ac} \, \partial_b H \, \partial_d b^{cd}
		+ b_a{}^c \, \partial_d b_b{}^d \, \partial_c H
		+ b_b{}^c \, \partial_d b_a{}^d \, \partial_c H
	\Big]
\notag \\
& \quad
- {1 \over 2} H^{-3} \, b_a{}^c b_b{}^d \, \partial_c \partial_d H
- {1 \over 2} H^{-2} \big( \delta_{ab} \delta^{cd} \, \partial_c H \partial_d H
		- 2 \partial_a H \partial_b H \big)
- {1 \over 2} H^{-1} \partial_a \partial_b H,
\\ \notag \\
{\mathcal K}_{\bar{a}\bar{b}}
&= {1 \over 2} H^{-4}
	\big( 2 \delta^c_{\bar{a}} \delta^d_{\bar{b}} \, \partial_c H \partial_d H
	- \delta_{\bar{a}\bar{b}} \delta^{cd} \, \partial_c H \partial_d H \big)
	- {1 \over 2} H^{-3} (\delta^c_{\bar{a}} \delta^d_{\bar{b}} \, \partial_c \partial_d H),
\\ \notag \\
{\mathcal K}_{a\bar{b}}
&= -{1 \over 2} H^{-4}
	\big( b_{a\bar{b}} \, \delta^{cd} \, \partial_c H \partial_d H - 2 \delta_{\bar{b}}^d \, b_a{}^c \, \partial_c H \partial_d H \big)
\notag \\
& \quad
	- {1 \over 4} H^{-3}
	\big( \partial_a H \, \partial_c b_{\bar{b}}{}^d
	+ \delta_{\bar{b}}^c \, \partial_d b_a{}^d \, \partial_c H
	+ 2 \delta_{\bar{b}}^d \, b_a{}^c \, \partial_d \partial_c H
	\big).
\end{align}
Using these explicit forms of the components, we can show that the second component of the projection $P_{MN} {}^{KL}$ (in
\eqref{eq:projection}) on the $\mathcal{K}$-tensor satisfies
$\eta_{MK} {\mathcal H}^{KP} {\mathcal K}_{PQ} {\mathcal H}^{QL} \eta_{LN} = {\mathcal K}_{MN}$.
This eventually leads to the fact that the second
equation in \eqref{eq:eom} holds.
Then we conclude that the ansatz \eqref{eq:DFT_monopole_ansatz}
becomes a solution to DFT provided the
harmonic function $H(y)$ satisfies the condition \eqref{eq:BPS_condition}.

It is worthwhile to convince ourselves by showing that the DFT ansatz
\eqref{eq:DFT_monopole_ansatz} indeed contains well-known supergravity
solutions. For example, by
assigning geometrical coordinates to a half of $X^M$ and
comparing the codimension-four ansatz
\eqref{eq:DFT_monopole_ansatz} with the parametrization of the
generalized metric \eqref{eq:generalized_metric}
\begin{align}
ds^2_{\text{DFT}}
&= (g_{\mu\nu}
	- B_{\mu\rho} g^{\rho\sigma} B_{\sigma\nu})
		dx^\mu dx^\nu
	+ 2 (B_{\mu\rho} g^{\rho\nu}) dx^\mu d\bar{x}_\nu
	+ g^{\mu\nu} d\bar{x}_\mu d\bar{x}_\nu,
\label{eq:KK_like_ansatz}
\end{align}
we can write down explicit solutions for conventional supergravity fields.
Choosing the geometrical coordinates $x^\mu = (x^m, y^i, y^9)$
$(i = 6,7,8)$,
we obtain the following solution
\begin{align}
ds^2
&=
\eta_{mn} dx^m dx^n + H (\delta_{ij} dy^i dy^j + (dy^9)^2),
\notag \\
B
&= 	(b_{ij} dy^i - b_{j9} dy^9) \wedge dy^j,
\notag \\
e^{2\phi}
&=	H,
\label{eq:NS5-brane}
\end{align}
where $ds^2 = g_{\mu \nu} dx^{\mu} dx^{\nu}$,
$H (y^i, y^9) = c + \frac{Q}{r^2}$, $r^2 = (y^i)^2 + (y^9)^2$;
$B$ is the Kalb-Ramond field, $\phi$ is the dilaton in
supergravity, and $c,Q$ are constants.
This is nothing but the NS5-brane solution in type II supergravity.
The condition \eqref{eq:BPS_condition} is just the one of the 1/2 BPS
condition for the NS5-brane solution \cite{Callan:1991dj}.
This seems plausible since DFT is reduced to the NSNS sector of
supergravity by imposing the strong condition.
Solutions to DFT should contain those for supergravity.

A remarkable fact about DFT is that, starting from a known
supergravity solution, one can obtain another solution by an $O(d,d)$
transformation. The $O(d,d)$ transformation changes the role of
the geometrical and winding coordinates.
For example, by choosing a different set of the coordinates
$x^\mu = (x^m, y^i, \bar{y}_9)$ as the geometrical ones, we obtain a new solution,
\begin{align}
ds^2
&=
\eta_{mn} dx^m dx^n + H^{-1} (d\bar{y}_9 + b_{i9} dy^i)^2 + H \delta_{ij} dy^i dy^j,
\notag \\
B
&= 	b_{ij} dy^i \wedge dy^j,
\notag \\
e^{2\phi}
&=	 {\rm const.}
\label{eq:KK-monopole_1winding}
\end{align}
Here, again the harmonic function $H = H(y^i,y^9)$ is given by the one in the
NS5-brane but now one of the coordinate $y^9$ is {\it not} the
geometrical but the T-dualized {\it winding} coordinate.
We therefore denote it as $y^9 = \tilde{y}_9$.

This solution is first obtained in \cite{Berman:2014jsa}.
When the harmonic function is smeared along the $\tilde{y}_9$-direction,
the solution \eqref{eq:KK-monopole_1winding} acquires an isometry and
the harmonic function becomes
$H(y^i, \tilde{y}_9) \to H(y^i) = c + \frac{Q'}{r'}$.
Here $Q'$ is a constant and $r^{\prime 2} = (y^i)^2$. Then the solution
becomes the KK-monopole. On the other hand, for the solution that is not
smeared, the KK-monopole has the winding coordinate dependence.
It is evident that the winding corrected KK-monopole solution
\eqref{eq:KK-monopole_1winding} is obtained by formally applying the Buscher rule
to the NS5-brane solution \eqref{eq:NS5-brane}.
Although the conventional T-duality transformations need isometries of
backgrounds \cite{Buscher:1987sk}, it is nevertheless possible to perform
a T-duality transformation without isometries \cite{delaOssa:1992vci}.
This is called the generalized T-duality \cite{Dabholkar:2005ve}.
A natural interpretation is based on string field theory in the toroidal
compactification \cite{Kugo:1992md}.
The physical interpretation of the winding coordinate dependence of the
KK-monopole was first discussed in \cite{Gregory:1997te}.
Later it is recognized that this is due to the string worldsheet instanton
effects \cite{Harvey:2005ab}. We will discuss this issue in the
subsequent sections.

It is possible to proceed the above procedure further.
Choosing yet another set of the geometrical coordinates
$x^\mu = (x^m, y^\alpha, \bar{y}_8, \bar{y}_9)$ and
comparing \eqref{eq:KK_like_ansatz} with \eqref{eq:DFT_monopole}, we obtain the
following solution:
\begin{align}
ds^2
&= \eta_{mn} \, dx^m dx^n
	+ H \delta_{\alpha\beta} \, dy^\alpha dy^\beta
\notag \\
& \quad
	+ {H \over H^2 + b_{89}^2} \big[ (d\bar{y}_9 + b_{\alpha 9} dy^\alpha)^2
		+ (d\bar{y}_8 + b_{\alpha 8} dy^\alpha)^2 \big],
\notag \\
B
&= b_{\alpha\beta} \, dy^\alpha \wedge dy^\beta
	- {b_{89} \over H^2 + b_{89}^2} \Big[
		(d\bar{y}_8 + b_{\alpha 8} dy^\alpha) \wedge
		(d\bar{y}_9 + b_{\beta 9} dy^\beta) \Big],
\notag \\
e^{2\phi}
&= {H \over H^2 + b_{89}^2},
\label{eq:Q5-brane_2winding}
\end{align}
where $\alpha = 6,7$.
Once again, the harmonic function
$H = H(y^\alpha, y^8, y^9)$
is given by the one in the NS5-brane
but two of the coordinates $(y^8 = \tilde{y}_8, y^9 = \tilde{y}_9)$  are not geometrical.
They are winding coordinates in this frame.
The solution \eqref{eq:Q5-brane_2winding} represents a five-brane
obtained by formal T-duality transformations.
This kind of brane is known as a $Q$-brane.
This becomes more evident when we perform the smearing procedure along the
$\tilde{y}_8$- and $\tilde{y}_9$-directions.
After the smearing, the solution \eqref{eq:Q5-brane_2winding}
is reduced to the exotic $5^2_2$-brane \cite{deBoer:2010ud}.
The $5^2_2$-brane does not have any winding coordinate dependence and
seems a geometric object.
However, one can show that the background geometry of the $5^2_2$-brane
inherits the non-trivial $O(2,2)$ monodromy.
In this sense, the $5^2_2$-brane is called the globally non-geometric object.
$B$ is a Kalb-Ramond field
given as the mixed symmetry field \cite{Bergshoeff:2011se}
on the $5^2_2$-brane background.
The winding corrections appear not only in the harmonic function
but also in the Kalb-Ramond field.
When only the $\tilde{y}_9$-direction
is smeared, the solution \eqref{eq:Q5-brane_2winding} is reduced to the
$5^2_2$-brane with one winding correction.
It was discussed that this modified $5^2_2$-brane with the one winding
coordinate dependence is indeed embedded into the codimension three
DFT monopole solution \cite{Bakhmatov:2016kfn}.

Now it is straightforward to proceed further.
By choosing
$x^\mu = (x^m, y^6, \bar{y}_{\hat{\imath}})$ $(\hat{\imath} = 7,8,9)$ as
the geometrical coordinate and comparing \eqref{eq:KK_like_ansatz} with \eqref{eq:DFT_monopole},
we obtain the following solution:
\begin{align}
ds^2
&= \eta_{mn} \, dx^m dx^n + H(dy^6)^2
\notag \\
& \quad
	+ K_2^{-1} \left[
		H^2 (d\bar{y}_{\hat{k}} + b_{6\hat{k}} dy^6)^2
		+ \left(
		\frac{1}{2} \varepsilon^{\hat{\imath}\hat{\jmath}\hat{k}} b_{\hat{\imath}\hat{\jmath}} (d\bar{y}_{\hat{k}} + b_{6\hat{k}} dy^6)
		\right)^2 \right],
\notag \\
B
&= - \frac{H b_{\hat{\imath}\hat{\jmath}}}{K_2} \, (d\bar{y}_{\hat{\imath}} + b_{6\hat{\imath}} dy^6) \wedge d\bar{y}_{\hat{\jmath}},
\notag \\
e^{2\phi} &= H K_2^{-1}, \notag \\
K_2 &\equiv H(H^2 + b_{89}^2 + b_{79}^2 + b_{78}^2).
\label{eq:R-brane}
\end{align}
The harmonic function in this solution is
$H = H(y^6, y^7, y^8, y^9)$ where $y^6$ is the geometric coordinates
and $y^7 = \tilde{y}_7, y^8 = \tilde{y}_8, y^9 = \tilde{y}_9$ are the winding coordinates in this duality frame.
The solution \eqref{eq:R-brane} is a five-brane obtained by formally
applying the Buscher rule to the $5^2_2$-brane.
This is known as the $R$-brane and conventionally denoted as the $5^3_2$-brane.
We remark that it is not possible to obtain a geometric five-brane in
this frame by smearing the winding directions $\tilde{y}_7, \tilde{y}_8, \tilde{y}_9$.
Even though the smearing works for the harmonic function,
the condition \eqref{eq:BPS_condition} implies that the solution becomes
trivial.
In order to keep the non-trivial structure of the fields $g_{\mu \nu},
B_{\mu \nu}$ and $\phi$, at least one of the winding coordinate should
be kept intact. Therefore the $R$-brane is a locally non-geometric object.

We are now in a position of the final possibility.
By choosing yet another different geometrical coordinates
$x^\mu 
= (x^m, \bar{y}_a)\ (a=6,7,8,9)$
and comparing \eqref{eq:KK_like_ansatz} with \eqref{eq:DFT_monopole},
we obtain the following five-brane solution:
\begin{align}
ds^2
&= \eta_{mn} \, dx^m dx^n
+ {H \over K_3} \left[ \left( H^2 + {1 \over 2} b_{cd} b^{cd} \right) \delta^{ab}
	- \delta_{cd} \, b^{ca} b^{db} \right] d\bar{y}_a d\bar{y}_b,
\notag \\
B
&= K_3^{-1} \left[ b^{cd} b_{ca} b_{db}
	- \left( H^2 + {1 \over 2} b_{cd} b^{cd} \right) b_{ab}
	\right]
d \bar{y}_a \wedge d \bar{y}_b,
\notag \\
e^{2\phi} &= H K_3^{-1},
\notag \\
K_3 &\equiv H^4 + {1 \over 2} H^2 \, b_{ab} b^{ab}
	+ \left( {1 \over 8} \varepsilon^{abcd} b_{ab} b_{cd} \right)^2.
\label{eq:542}
\end{align}
This is denoted as the $5^4_2$-brane solution.
The harmonic function in this $5^4_2$-brane solution is
$H = H (\tilde{y}_6, \tilde{y}_7, \tilde{y}_8, \tilde{y}_9)$
where all the coordinates represent the winding directions.
If we smear all these directions, the solution becomes a codimension zero,
space-filling five-brane in six dimensions.

\subsubsection{Codimension three}
We next study five-brane solutions of codimension three.
These solutions are obtained by smearing, for example, the $y^9$-direction in
the five-branes of codimension four.
Here, considering the relation between $b_{ab}$ and harmonic function \eqref{eq:BPS_condition},
only $b_{i9} = A_i$ ($i=6,7,8$) is non-zero component.
In this way, the ansatz \eqref{eq:DFT_monopole_ansatz}
for the codimension four five-brane is reduced to the following form:
\begin{align}
ds^2_{\text{DFT}}
&= H(1 + H^{-2} A^2) (dy^9)^2 + H^{-1} (d\bar{y}_9)^2
	+ 2H^{-1} A_i [ dy^i \, d\bar{y}_9 - \delta^{ij} \, d\bar{y}_j \, dy^9]
\notag \\
&\quad
	+ H(\delta_{ij} + H^{-2} A_i A_j) \, dy^i dy^j
	+ H^{-1} \delta^{ij} \, d\bar{y}_i d\bar{y}_j
\notag \\
& \quad
	+ \eta_{mn} \, dx^m dx^n + \eta^{mn} \, d\bar{x}_m d\bar{x}_n.
\label{eq:codim_3_DFT_monopole}
\end{align}
This is the DFT monopole ansatz discussed first in \cite{Berman:2014jsa}.
It is straightforward to show that the ansatz
\eqref{eq:codim_3_DFT_monopole} satisfies the DFT equations of motion
\eqref{eq:eom} provided that the $b_{ab}$ satisfies the relation
${\text{rot}\,} \vec{A} = {\text{grad}\,} H$.
Following the same steps in the case of the codimension four, we obtain
branes of codimension three.
\begin{itemize}

\item
H-monopole:
\begin{align}
ds^2
&=
\eta_{mn} \, dx^m dx^n
+ H (\delta_{ij} dy^i dy^j + (dy^9)^2),
\notag \\
B
&= 	- A_j \, dy^9 \wedge dy^j,
\notag \\
e^{2\phi}
&=	H.
\label{eq:H-monopole}
\end{align}
This is the H-monopole obtained by smearing the $y^9$-direction of the
     NS5-brane solution \eqref{eq:NS5-brane}.
The harmonic function is given by $H(r') = c + \frac{Q'}{r'}$, where
     $c,Q'$ are constants and $r^{\prime 2} = (y^i)^2$.

\item KK-monopole:
\begin{align}
ds^2
&=
\eta_{mn} \, dx^m dx^n
+ H^{-1} (d\bar{y}_9 + A_i dy^i)^2 + H \delta_{ij} \, dy^i dy^j,
\notag \\
B
&= 	0,
\notag \\
e^{2\phi} &= \text{const.}
\label{eq:KK-monopole}
\end{align}
This is the KK-monopole obtained by smearing the winding direction
      $\tilde{y}_9$ of the solution \eqref{eq:KK-monopole_1winding}.
The harmonic function is the same with the one for the H-monopole.

\item $5^2_2$-brane with one-winding dependence:
\begin{align}
ds^2
&= \eta_{mn} \, dx^m dx^n
	+ H \delta_{\alpha\beta} \, dy^\alpha dy^\beta
	+ {H \over H^2 + A_8^2} \big[ (d\bar{y}_9 + A_\alpha dy^\alpha)^2
		+ (d\bar{y}_8)^2 \big],
\notag \\
B
&= - {A_8 \over H^2 + A_8^2} \Big[
		d\bar{y}_8  \wedge
		(d\bar{y}_9 + A_\beta dy^\beta) \Big],
\notag \\
e^{2\phi} &= {H \over H^2 + A_8^2}.
\label{eq:Q5-brane_1winding}
\end{align}
This is obtained by smearing one of the two winding directions
$\tilde{y}_9$ of the solution \eqref{eq:Q5-brane_2winding}.
The harmonic function is given by $H(r') = c + \frac{Q'}{r'}$, where
      $r^{\prime 2} = (y^6)^2 + (y^7)^2 + (\tilde{y}^8)^2$.

\item $5^3_2$-brane with two-winding dependence:
\begin{align}
ds^2
&= \eta_{mn} dx^m dx^n
+ H (dy^6)^2
\notag \\
& \quad
	 + {H^2 \over K'_2} [ d\bar{y}_7^2 + d\bar{y}_8^2 + (d\bar{y}_9 - A_6 dy^6)^2]
	+ (K'_2)^{-1} \left[ A_8 \, d\bar{y}_7 - A_7 \, d\bar{y}_8 \right]^2,
\notag \\
B
&= - H (K'_2)^{-1} (A_7 \, d\bar{y}_7 + A_8 \, d\bar{y}_8) \wedge (d\bar{y}_9 + A_6 \, dy^6),
\notag \\
e^{2\phi} &= H (K'_2)^{-1},
\notag \\
K'_2 &\equiv H(H^2 + A_7^2 + A_8^2).
\label{eq:codim3_R-brane}
\end{align}
This is obtained by smearing one of the three winding directions $\tilde{y}_9$
of the $R$-brane solution \eqref{eq:R-brane}.
The harmonic function is given by $H(r') = c + \frac{Q'}{r'}$, where
      $r^{\prime 2} = (y^6)^2 + (\tilde{y}^7)^2 + (\tilde{y}^8)^2$.

\item $5^4_2$-brane with three-winding dependence:
\begin{align}
ds^2
&= \eta_{mn} \, dx^m dx^n
+ H^{-1} (\delta^{ij} \, d\bar{y}_i d\bar{y}_j + d\bar{y}_9^2)
	- {H \over K'_3} \big[ (A^i d\bar{y}_i)^2 + A^2 \, d\bar{y}_9^2
 \big],
\notag \\
B
&= - H^2 (K'_3)^{-1} A_i d\bar{y}_i \wedge d\bar{y}_9,
\notag \\
e^{2\phi} &= H (K'_3)^{-1},
\notag \\
K'_3 &\equiv H^2 (H^2 + A_6^2 + A_7^2 + A_8^2).
\label{eq:codim3_542}
\end{align}
This is obtained by smearing one of the four winding directions $\tilde{y}_9$
of the space-filling brane \eqref{eq:542}.
The harmonic function is given by $H(r') = c + \frac{Q'}{r'}$, where
      $r^{\prime 2} = (\tilde{y}^6)^2 + (\tilde{y}^7)^2 + (\tilde{y}^8)^2$.
\end{itemize}
The solutions \eqref{eq:H-monopole} and \eqref{eq:KK-monopole} are first
found in \cite{Berman:2014jsa} while \eqref{eq:Q5-brane_1winding} is
found in \cite{Bakhmatov:2016kfn}.
The others \eqref{eq:codim3_R-brane} and \eqref{eq:codim3_542} are new solutions.

\subsubsection{Codimension two}
We then show five-brane solutions of codimension two.
To this end, it is convenient to employ the cylindrical coordinates
$y^6 = \rho \cos \theta$, $y^7 = \rho \sin \theta$, $y^8$.
Codimension two solutions are obtained by smearing the
one more transverse direction in the harmonic function.
This results in the logarithmic form of the harmonic function.
It is possible to make only $A_8$ non-zero by gauge transformation.
We fix $A_8$ to $A_8 = \sigma\theta$, where $\sigma$ is a constant.
Thus, the DFT monopole ansatz \eqref{eq:codim_3_DFT_monopole}
becomes the ``codimension-two DFT monopole'':
\begin{align}
ds^2_{\text{DFT}}
&= H(1 + H^{-2} A_8^2) (dy^9)^2 + H^{-1} d\bar{y}_9^2
	+ 2H^{-1} A_8 [ dy^8 \, d\bar{y}_9 - d\bar{y}_8 \, dy^9]
\notag \\
&\quad
	+ H \delta_{ij} \, dy^i dy^j + H^{-1} A_8^2 (dy^8)^2
	+ H^{-1} \delta^{ij} \, d\bar{y}_i d\bar{y}_j
\notag \\
& \quad
	+ \eta_{mn} \, dx^m dx^n + \eta^{mn} \, d\bar{x}_m d\bar{x}_n.
\label{eq:codim2DFTm}
\end{align}
By the same procedure discussed above,
 various five-brane solutions with harmonic functions of codimension-two
 are obtained.
\begin{itemize}

\item Smeared H-monopole:
\begin{align}
ds^2
&=
\eta_{mn} \, dx^m dx^n
+ H \delta_{ab} \, dy^a dy^b,
\notag \\
B
&= 	- A_8 \, dy^9 \wedge dy^8,
\notag \\
e^{2\phi}
&=	H.
\label{eq:smeared_H-monopole}
\end{align}
Here the harmonic function is
$H = h_0 + q \log \frac{\mu}{\rho}$ where $h_0, q, \mu$ are constants and
$\rho^2 = (y^6)^2 + (y^7)^2$.

\item Smeared KK-monopole (KK-vortex)
\cite{Onemli:2000kb}:
\begin{align}
ds^2
&=
\eta_{mn} \, dx^m dx^n
+ H^{-1} (d\bar{y}_9 + A_8 \, dy^8)^2 + H \delta_{ij} \, dy^i dy^j,
\notag \\
B
&= 	0,
\notag \\
e^{2\phi} &= \text{const.}
\end{align}
The harmonic function is the one for the smeared H-monopole.

\item $5^2_2$-brane:
\begin{align}
ds^2
&= \eta_{mn} \, dx^m dx^n
	+ H \delta_{\alpha\beta} \, dy^\alpha dy^\beta
	+ {H \over H^2 + A_8^2} \big[ d\bar{y}_9^2 + d\bar{y}_8^2 \big],
\notag \\
B
&= - \frac{A_8}{H^2 + A_8^2} \, d\bar{y}_8 \wedge d\bar{y}_9,
\notag \\
e^{2\phi} &= \frac{H}{H^2 + A_8^2}.
\end{align}
The harmonic function is again the one for the smeared H-monopole.

\item $5^3_2$-brane with one-winding dependence:
\begin{align}
ds^2
&= \eta_{mn} \, dx^m dx^n
+ H (dy^6)^2
	+ H^{-1} d\bar{y}_7^2
	+ \frac{H}{H^2 + A_8^2}  [d\bar{y}_8^2 + d\bar{y}_9^2],
\notag \\
B
&= - \frac{A_8}{H^2 + A_8^2} \, d\bar{y}_8 \wedge d\bar{y}_9,
\notag \\
e^{2\phi} &= \frac{1}{H^2 + A_8^2}.
\label{eq:532_1winding}
\end{align}
Here the harmonic function is
$H = h_0 + q \log \frac{\mu}{\rho}$
where $\rho^2 = (y^6)^2 + (\tilde{y}_7)^2$.

\item $5^4_2$-brane with two-winding dependence:
\begin{align}
ds^2
&= \eta_{mn} \, dx^m dx^n
+ H^{-1} \delta^{ab} \, d\bar{y}_a d\bar{y}_b
	- \frac{H^{-1}}{H^2 + A_8^2}
\big[ (A_8 \, d\bar{y}_8)^2 + A_8^2 \, d\bar{y}_9^2 \big],
\notag \\
B
&= - \frac{A_8}{H^2 + A_8^2} \, d\bar{y}_8 \wedge d\bar{y}_9,
\notag \\
e^{2\phi} &= \frac{H^{-1}}{H^2 + A_8^2}.
\label{eq:542_2winding}
\end{align}
Here the harmonic function is
$H = h_0 + q \log \frac{\mu}{\rho}$
where $\rho^2 = (\tilde{y}_6)^2 + (\tilde{y}_7)^2$.
\end{itemize}
The last two \eqref{eq:532_1winding} and \eqref{eq:542_2winding} are new solutions.

\subsubsection{Codimensions one and zero}
In order to obtain branes of codimension one, we perform the smearing
along, for example, the $y^7$-direction.
Then the harmonic function becomes a linear function of the remaining
one direction $y^6$; $H(y^6) = h_0 - m |y^6|$, where $h_0$ and $m$ are
constants.
However, since the condition \eqref{eq:BPS_condition} implies
$\partial_6 H(y^6) = \partial_7 A_8 = -m$, the gauge field $A_8$ should
be a constant and $m = 0$
\footnote{
\label{ft:cd1}
We note that it is possible to keep $m$ nonzero when one allows the non-constant gauge field $A_8 = -m y^7$.
In this case, the solution is classified into the codimension two objects in our convention.
Even though the harmonic function is a linear function of $y^6$ or $\tilde{y}_6$, each brane solution takes the same form as in
\eqref{eq:smeared_H-monopole}-\eqref{eq:542_2winding}.
The solution \eqref{eq:532_1winding} with $H = h_0 - m |y^6|$, $A_8 = - m \tilde{y}_7$ is usually referred to as the $R$-brane in the literature.
}.
Because of this reduction of the harmonic function and the gauge field, the winding coordinate corrections to the $5^4_2$-brane disappear, and the solution seems to be simple without any non-trivial structures. However,
one can still consider the ansatz \eqref{eq:KK_like_ansatz} for branes
of codimension one.

Furthermore, if we smear the harmonic function also in the
remaining direction,
the harmonic function becomes constant.
In the following,
we fix the gauge field $A_8 = 0$.
The DFT monopole represents the flat solution in the doubled space
\begin{align}
ds^2_{\text{DFT}}
&= H \delta_{ab} \, dy^a dy^b
+ H^{-1} \delta^{ab} \, d\bar{y}_a d\bar{y}_b
+ \eta_{mn} \, dx^m dx^n + \eta^{mn} \, d\bar{x}_m d\bar{x}_n.
\label{eq:codim0DFTm}
\end{align}
In summary, we have classified all the five-brane solutions in the T-duality
chains starting from the NS5-brane, H-monopole, smeared H-monopole,
and doubly smeared H-monopole. We have also obtained the T-duality chain
starting from the space-filling five-brane (triply smeared H-monopole).
They are listed in Table \ref{tab:various_fivebrane}.
We stress that five-branes with multiple winding corrections naturally
appear in the DFT solutions. We will discuss a physical interpretation
of these multiple winding corrections in section \ref{sect:instanton}.

\begin{table}[tb]
\begin{center}
\begin{tabular}{c  ccccc} \hline
 & $5_2$ & $5^1_2$ & $5^2_2$ & $5^3_2$ & $5^4_2$ \vspace{1pt} \\ \hline \hline
$\text{codim} = 4$ & NS5 & KKM + $w^1$ & $5^2_2 + w^2$ & $5^3_2 + w^3$ & $5^4_2 + w^4$ \\
$\text{codim} = 3$ & HM & KKM & $5^2_2 + w^1$ & $5^3_2 + w^2$ & $5^4_2 + w^3$ \\
$\text{codim} = 2$ & sHM & sKKM & $5^2_2$ & $5^3_2 + w^1$ & $5^4_2 + w^2$ \\
$\text{codim} = 1$ & dsHM & dsKKM & s$5^2_2$ & $5^3_2$ & $5^4_2$ \\
$\text{codim} = 0$ & tsHM & tsKKM & ds$5^2_2$ & s$5^3_2$ & $5^4_2$ \\ \hline
\end{tabular}
\caption{
Various five-brane solutions obtained in this section.
The notation in this table is as follows:
``HM'' refers to as the H-monopole, and ``KKM'' refers to as the KK-monopole:
``s'', ``ds'' and ``ts'' are abbreviations of ``smeared'', ``doubly smeared'' and ``triply smeared'', respectively.
Also, the $n$-winding dependence is written as $w^n$.
For the notation of the $b^c_n$-brane, see \cite{deBoer:2010ud}.
}
\label{tab:various_fivebrane}
\end{center}
\end{table}

\subsection{Wave solutions}
In this section, we study other extended objects in the NSNS sector, namely, the
F-string and wave.
It was found in \cite{Berkeley:2014nza} that the supergravity
solutions of the F-string and the wave are embedded in the DFT solution
known as the DFT wave.
The DFT wave solution is given by
\begin{align}
ds^2_{\text{DFT}}
&= {\mathcal H}_{MN} \, dX^M dX^N
\notag \\
&= (H-2) dt^2 + (2-H) dz^2 + 2(1-H) (dt \, d\bar{z} + d\bar{t} \, dz)
\notag \\
& \quad
	 - H d\bar{t}^{\, 2} + H d\bar{z}^2
	+ \delta_{ij} \, dy^i dy^j + \delta^{ij} \, d\bar{y}_i d\bar{y}_j,
\notag \\
d &= {\rm const}.
\label{eq:DFT_wave}
\end{align}
Here the generalized coordinates $M=1,\ldots,20$ are defined by
\begin{align}
X^M =(t, z, y^i; \bar{t}, \bar{z}, \bar{y}_i),
\qquad i=1,\ldots,8,
\end{align}
where $t,z$, $y^i$ are the worldvolume and the transverse directions of
the F-string and the wave while the bared coordinates are their dual pairs.
The DFT wave solution is governed by the harmonic function $H = H(y)$.
We first consider objects of codimension eight by bearing in mind the F-string
solution. Then the harmonic function is given by
\begin{align}
H = c + \frac{h}{r^6},
\qquad r^2 = \delta_{ij} \, y^i y^j,
\label{eq:codim8}
\end{align}
where $c,h$ are an arbitrary constant.

We look for a conventional object in ten dimensions.
By comparing the KK-like ansatz \eqref{eq:KK_like_ansatz}
with \eqref{eq:DFT_wave}, we obtain the metric, the Kalb-Ramond field
and the dilaton in the conventional supergravity.
If we choose the geometrical coordinates $x^\mu = (t,z,y^i)$,
we obtain the F-string extended along the $z$-direction with the
Kalb-Ramond field and the dilaton $\phi$
\begin{align}
ds^2
&=	- H^{-1} (dt^2 - dz^2) + \delta_{ij} \, dy^i dy^j,
\notag \\
B
&=	- (H^{-1} -1) dt \wedge dz,
\notag \\
e^{2\phi} &= H^{-1}.
\label{eq:F-string_std}
\end{align}
On the other hand, if we choose the coordinates
$x^\mu = (t,\bar{z},y^m)$, we obtain the pp-wave solution:
\begin{align}
ds^2
&=	(H-2) dt^2 + H d\bar{z}^2 + 2 (H-1) dt \, d\bar{z} + \delta_{ij} \, dy^i dy^j,
\notag \\
B &= 0,
\notag \\
e^{2\phi} &= \text{const.}
\label{eq:pp-wave}
\end{align}
This is a conceivable result since the F-string and the wave are related
by the T-duality transformation.
These solutions were discussed in \cite{Berkeley:2014nza}.
Now we follow the same discussion of the five-branes case.
By choosing the geometrical coordinates
$x^\mu = (t,z,y^{i'},\bar{y}_8)$ ($i' = 1, \ldots, 7$),
we can further T-dualize the solution. The resulting solution is
\begin{align}
ds^2
&= 	- H^{-1} (dt^2 - dz^2) + d\bar{y}_8^2 + \delta_{i'j'} \, dy^{i'} dy^{j'},
\notag \\
B
&=	- (H^{-1} -1) dt \wedge dz,
\notag \\
e^{2\phi} &= H^{-1}.
\label{eq:F-string_winding}
\end{align}
The only difference between the solutions \eqref{eq:F-string_winding} and
\eqref{eq:F-string_std} is that $y^8$ and $\bar{y}_8$ are interchanged.
Again, the harmonic function is given by \eqref{eq:codim8} but now one
of the transverse direction has been changed to winding direction
$y^8 = \tilde{y}_8$, namely
$r^2 = (y^{i'})^2 + (\tilde{y}_8)^2$.
Since the solution \eqref{eq:F-string_winding} is obtained by formally
applying the Buscher rule to the F-string solution, it can be
interpreted as an F-string localized in the one-winding direction\footnote{
The F-string localized in the winding space is discussed in
\cite{Blair:2016xnn}.}.
Selecting more $\bar{y}_i$ as geometrical coordinates, we can obtain
other F-string solutions that depend on more winding coordinates.
Even if the F-string has more winding coordinate dependence, the form of the solution is the same as~\eqref{eq:F-string_winding}.

On the other hand, choosing the geometrical coordinates
$x^\mu = (t, \bar{z}, y^{i'}, \bar{y}_8)$, we obtain another wave solution
\begin{align}
ds^2
&=	(H-2) dt^2 + H d\bar{z}^2 + 2 (H-1) dt \, d\bar{z}
	+ \delta_{i'j'} \, dy^{i'} dy^{j'} + d\bar{y}_8^2,
\notag \\
B &= 0,
\notag \\
e^{2\phi} &= \text{const.}
\end{align}
The form of this solution is similar to~\eqref{eq:pp-wave},
but the harmonic function depends on the winding coordinate $y^8 = \tilde{y}_8$.
As in the case of the F-string,
wave solutions that depend on more winding coordinates are obtained by
selecting more $\bar{y}_i$ as geometrical coordinates.
Since the winding dependence is obtained by
T-duality transformation along the transverse directions, the winding
corrected F-string and the pp-wave are defined up to the eight winding coordinates.

\section{Manifesting the winding corrections} \label{sect:resum}

In this section,
we make the following issue manifest:
the winding coordinate dependence discussed in the previous section is
the ``non-geometric'' corrections to the ordinary supergravity solution.
The analysis is based on the rewriting of the harmonic function in the
discrete mode expansions.
In the following, we examine the winding corrections step-by-step in the
T-duality orbit for the five-branes of codimension four.
Although, the DFT solutions discussed in the previous section do not
require compactifications along the T-duality directions, we understand
that this is necessary for physical interpretations.
We therefore assume that the directions where the T-duality transformations
have been performed are compactified on an $N$-dimensional torus $T^N$.

\subsection{H- and KK-monopoles}
We start from the NS5-brane solution in ten-dimensional type II
supergravities.
The solution is
\begin{align}
ds^2 = dx^2_{012345} + H (r) dx^2_{6789}, \quad
e^{2\phi} = H (r), \quad H_{\mu \nu \rho}
= \varepsilon_{\mu \nu \rho \sigma} \partial^{\sigma} H (r),
\end{align}
where the world-volume is extended along the $x^{0,1,2,3,4,5}$-directions
while the transverse directions are denoted by $x^{\mu}, \ (\mu = 6,7,8,9)$.
The 1/2 BPS equation of supergravity implies that $H$ is a harmonic
function in the transverse space: $\Box H = 0$, where $\Box = \sum_{\mu
= 6}^9 (\partial_{\mu})^2$.
The harmonic function of the NS5-brane solution is therefore
\begin{align}
H (r) = c + \frac{Q}{r^2}, \qquad
r^2 = (x^6)^2 + (x^7)^2 + (x^8)^2 + (x^9)^2,
\end{align}
where $c$ and $Q$ are constants.
Since the (non)geometries are completely determined by the harmonic
function $H$, we focus on the structure of $H$ in the following.

Now we compactify the $x^9$-direction to
a circle $S^1$ with its radius $R_9$.
Then the solution becomes a periodic array of the NS5-branes.
The harmonic function becomes
\begin{align}
H (r',x^9) = c +  \sum_{n = - \infty}^{\infty} \frac{Q}{(r')^2 + (x^9 - 2
 \pi R_9 n)^2}, \qquad (r')^2 =  (x^6)^2 + (x^7)^2 + (x^8)^2.
\label{eq:codim4}
\end{align}
By using the Poisson resummation formula\footnote{
In use of this formula, we need the absolute convergences in both sides
of the equation.
When the infinite summation is divergent, we need to justify the formula
by subtracting divergent pieces in the summation.
As we will see, divergent parts come from the ``zero-winding'' sectors
in codimension less than three cases.
}
\begin{align}
\sum_{n=-\infty}^{\infty} f (2 \pi n) = \frac{1}{2\pi}
 \sum_{n=-\infty}^{\infty} \int^{\infty}_{-\infty} \! dt \ f(t) e^{int},
\label{eq:Poisson}
\end{align}
we find
\begin{align}
H (r',x^9)
 =& \
 c +
 \frac{Q}{2 R_9} \frac{1}{r'}
 \left[
 1
 + \sum_{n=1}^{\infty} e^{\left( i \frac{x^9}{R_9} - \frac{r'}{R_9}
  \right) n}
 + \sum_{n=1}^{\infty} e^{\left( - i \frac{x^9}{R_9} - \frac{r'}{R_9}
  \right) n}
 \right]
\label{eq:codim4_3}
\\
=& \
c +
\frac{Q}{2 R_9} \frac{1}{r'}
\frac{\sinh (r' R_9^{-1})}{\cosh (r' R_9^{-1}) - \cos (x^9 R_9^{-1})}.
\label{eq:codim4_4}
\end{align}
This is the well-known harmonic function for a periodic array of five-branes.
In order to obtain
the H-monopole of codimension three, we consider the small radius limit $R_9 \to 0$.
In this limit, the harmonic function is reduced to
\begin{align}
H (r') = c + \frac{Q'}{r'},
\label{eq:codim3}
\end{align}
where we have kept $Q' = \frac{Q}{2R_9}$ finite.
This procedure is nothing but the smearing.
The smearing along the $x^9$-direction introduces the $U(1)$ isometry.
In this small $R_9$ limit, one can perform the T-duality transformation
of the H-monopole solution by the famous Buscher rule \cite{Buscher:1987sk}.
Then the resulting solution becomes the KK-monopole whose geometry is
realized as the Taub-NUT space where $B = \phi = 0$.
At infinity, the space is topologically $S^3$ which is given as
the Hopf fibration of $S^1$ over $S^2$.
The metric is governed by the harmonic function \eqref{eq:codim3}.

In the smearing procedure, one observes that the relevant term comes
from the $n = 0$ sector in \eqref{eq:codim4_3}
while the $n \neq 0$ KK-modes break the isometry.
From the perspective of strings propagating on the H-monopole
background, these $n\not=0$ terms are understood as worldsheet instanton
corrections \cite{Tong:2002rq}.

Let us consider the $n \not= 0$ corrections from the
T-dualized, namely, the KK-monopole side.
The worldsheet instanton calculus based on the GLSM technique was
performed in \cite{Harvey:2005ab}.
The worldsheet instantons in the Taub-NUT space introduces the
corrections to the harmonic function of the KK-monopole.
The result is
\begin{align}
H (r') \ \longrightarrow \ H (r', \tilde{x}_9) = c + \frac{Q'}{r'}
\left[
1 + \sum_{n \not= 0} \exp
\left(
i n \frac{\tilde{x}_9}{\tilde{R}_9} - r' \left| \frac{n}{\tilde{R}_9} \right|
\right)
\right].
\label{eq:mod_codim3}
\end{align}
This is precisely the expression \eqref{eq:codim4_3} but $x^9$ and $R_9$
are replaced by $\tilde{x}_9$ and $\tilde{R}_9$,
respectively\footnote{We should notice that the meaning of the symbol
tilde is different from that in
\cite{Kimura:2013zva}.
Throughout this paper we refer to the winding coordinate as $\tilde{x}_m$ in each configuration. This implies that $\tilde{x}_9$ in the KK-monopole system \eqref{eq:mod_codim3} represents the winding coordinate, which is nothing but the geometrical coordinate $x^9$ in the NS5-brane configuration \eqref{eq:codim4} and \eqref{eq:codim4_3}.
This seems confusing for readers. However, we emphasize the winding
corrections in each configuration in terms of tilde symbols. In the same
reason, we also refer to the radius of the dual winding coordinate
$\tilde{x}_i$ as $\tilde{R}_i$. This is related to the one of the
original coordinate in such a way as $\tilde{R}_i = \alpha'/R_i$.}.
Namely, the string worldsheet instanton corrections to the KK-monopole
geometry are identified with the winding coordinate corrections \cite{Harvey:2005ab}.

At first sight, this is a little bit puzzling.
From the KK-monopole viewpoint, the instanton corrections seem to break the isometry
in the {\it H-monopole side}, not the isometry in the Taub-NUT space.
However, this kind of the winding coordinate dependence of stringy geometry
is discussed in long ago \cite{Gregory:1997te} and it is a conceivable
result according to the correct counting of the zero-modes.
In the two-dimensional GLSM language, the winding coordinate dependence originates from
the total derivative term $\varepsilon^{ab} \partial_a (\tilde{X}_9A_b)$ \cite{Harvey:2005ab}.
Here $\tilde{X}_9$ is the scalar field that represents the $\tilde{x}_9$
coordinate and $A_a \ (a=0,1)$ is the $U(1)$ gauge field.
In the IR-limit of the GLSM, the two-dimensional field theory becomes
the non-linear sigma model (NLSM) probing the Taub-NUT space.
In this limit, $A_a$ is reduced to a non-dynamical field and should be integrated out.
This process makes the the corresponding term in the NLSM be a self-dual exact Kalb-Ramond field
$B = - \tilde{X}_9 d \Lambda$ in the Taub-NUT space. This $\tilde{X}_9$
is nothing but the missing dyonic zero-modes associated with the
NS5-brane position in the $x^9$-direction \cite{Sen:1997zb}. This
observation may justify the winding coordinate dependence of the geometry.

We point out  that the instanton corrected KK-monopole geometry based on
the modified harmonic function \eqref{eq:mod_codim3} does not satisfy the
supergravity equation of motion.
This is a reflection of the following fact.
The instanton number $n$ in \eqref{eq:mod_codim3} is just the string
winding number which is T-dual to the KK-mode number $n$ in
\eqref{eq:codim4_4}. As we have discussed above, the isometry-breaking
terms ({\it i.e.} the non-zero KK-modes $n \not=0$) in
\eqref{eq:codim4_4} becomes relevant when $\tilde{R}_9$ is large.
In the dual picture, the winding corrections become relevant when the radius of the $S^1$
fiber is small and the energy cost of strings wrapping on $S^1$ is economical.
Then, the string winding modes becomes lighter and lighter in the small
dual radius limit $\tilde{R}_9 \to 0$.
This substantially leads to the introduction of extra
massless fields to supergravity and they modify the equation of motion.
One finds that the harmonic function appeared in the DFT solution
\eqref{eq:KK-monopole_1winding} is just that we have shown in \eqref{eq:mod_codim3}.
It is worthwhile to emphasize that
the generalized T-duality without isometry leads to the string
worldsheet instanton corrections and provides an evidence of the stringy
realization of space-time.
In this process, the KK-modes are transformed to the
instanton corrections in the T-dual side.
We stress that this procedure is naturally incorporated in the DFT framework.

\subsection{$Q$-brane}
We proceed further T-duality in the KK-monopole configuration.
We begin with the same harmonic function as \eqref{eq:codim4} whilst the
variables $x^9$ and $R_9$ are replaced by $\tilde{x}_9$ and
$\tilde{R}_9$ due to the T-duality transformation from the NS5-brane
background.
In order to introduce an extra isometry, we compactify a transverse direction
(for instance, the $x^8$-direction)
to a circle $S^1$ and consider a periodic array of the winding
corrected KK-monopole.
Then the harmonic function becomes
\begin{align}
H (\rho,x^8, \tilde{x}_9) = c + \sum_{m=-\infty}^{\infty} \sum_{n=-\infty}^{\infty}
\frac{Q}{\rho^2 + (x^8 - 2 \pi R_8 m)^2 + (\tilde{x}_9 - 2 \pi \tilde{R}_9 n)^2},
\label{eq:codim3_1}
\end{align}
where $\rho^2 =  (x^6)^2 + (x^7)^2$.
We again notice that the variable $\tilde{x}_9$ represents the winding
coordinate in the KK-monopole configuration, which is the
``geometrical'' coordinate in the NS5-brane background.
Again, by using the Poisson resummation formula \eqref{eq:Poisson}, we find
\begin{align}
H(\rho,x^8, \tilde{x}_9) = c + \frac{Q}{2 \pi R_8 \tilde{R}_9}
\sum_{m=-\infty}^{\infty} \sum_{n=-\infty}^{\infty}
e^{im \frac{x^8}{R_8}} e^{in \frac{\tilde{x}_9}{\tilde{R}_9}} K_0
\left(
\rho
\sqrt{
\left(\frac{m}{R_8}\right)^2 +
\left(\frac{n}{\tilde{R}_9}\right)^2
}
\right),
\end{align}
where $K_0 (x)$ is the modified Bessel function of the second kind.
Note that \eqref{eq:codim3_1} is
a 
divergent series.
The divergence comes from $K_0(x=0)$.
This corresponds to the zero-mode sector with $n=m=0$.
In order to perform the Poisson resummation, we need to regularize the
expression \eqref{eq:codim3_1}.
This is easily done by subtracting the divergent piece from the
sum as in \cite{Lust:2017jox}. To see the structure of the divergent part,
we separate out the $n=m=0$
sector and use the asymptotic behavior of the modified Bessel function
around $x \sim 0$,
\begin{align}
K_0 (x) \sim (-\gamma + \log 2) - \log x + \frac{1}{4} (1 - \gamma +
 \log 2) x^2 - \frac{1}{4} x^2 \log x + \cdots,
\end{align}
where $\gamma = 0.577$ is the Euler's gamma constant.
Then we evaluate the zero-winding sector as
\begin{align}
\lim_{n,m \to 0}
e^{im \frac{x^8}{R_8}} e^{in \frac{\tilde{x}_9}{\tilde{R}_9}} K_0
\left(
\rho
\sqrt{
\left(\frac{m}{R_8}\right)^2 +
\left(\frac{n}{\tilde{R}_9}\right)^2
}
\right)
=
 \log \frac{\mu'}{\rho}
 - \lim_{n,m \to 0} \log \sqrt{
 \left(\frac{m}{R_8}\right)^2 +
 \left(\frac{n}{\tilde{R}_9}\right)^2
 }
+ \cdots,
\end{align}
where $\mu' = 2 e^{-\gamma}$ is a constant and the terms in
$\cdots$ vanish in the limit $m,n \to 0$.
It is obvious that the second term gives a logarithmic divergence.
The regulator of the form $\frac{Q}{2\pi R_8 \tilde{R}_9} \sum_{n > 0} \frac{1}{n}$
exactly cancels this divergence.
By subtracting this divergent piece from the right hand side of
\eqref{eq:codim3_1}, we justify the Poisson resummation process.
Performing the formal T-duality transformation along the $x^8$-direction
by the replacement $x^8$ and $R_8$ with $\tilde{x}_8$ and $\tilde{R}_8$
respectively, we obtain
\begin{align}
H(\rho,\tilde{x}_8,\tilde{x}_9) =& \ \frac{Q}{2 \pi \tilde{R}_8 \tilde{R}_9}
\left[
\log \frac{\mu}{\rho}
+
\sum_{n,m \not= (0,0)}
e^{i m \frac{\tilde{x}_8}{\tilde{R}_8}} e^{i n \frac{\tilde{x}_9}{\tilde{R}_9}}
K_0
\left(
\rho
\sqrt{
\left(\frac{m}{\tilde{R}_8}\right)^2 +
\left(\frac{n}{\tilde{R}_9}\right)^2
}
\right)
\right],
\label{eq:codim3_3}
\end{align}
where $\mu = 2 e^{- \gamma \frac{2 \pi \tilde{R}_8 \tilde{R}_9 c}{Q} }$.
This is the harmonic function appeared in the DFT solution
\eqref{eq:Q5-brane_2winding}.
Again, if we consider the smearing by the limit
$\tilde{R}_8, \tilde{R}_9 \to 0$ with $Q/2\pi \tilde{R}_8 \tilde{R}_9$
fixed, the $n,m \not= 0$ terms in \eqref{eq:codim3_3} vanish and
$H$ becomes a harmonic function for branes of codimension two.
The logarithmic behavior of the harmonic function implies that the
$5^2_2$-brane is not well-defined as a stand-alone object.
Indeed, the logarithmic part appears in the near brane limit of a globally
well-defined solution and the constant $\mu$ in \eqref{eq:codim2_reg}
is recognized as a cutoff where other duality branes become relevant
\cite{Kikuchi:2012za}.

On the other hand, if we take the limit
$\tilde{R}_8 \to \infty$ with $Q/2\pi \tilde{R}_8$
and $\tilde{R}_9$ fixed, we find
\begin{align}
H(\rho,\tilde{x}_9) = \frac{Q}{2 \pi \tilde{R}_8 \tilde{R}_9}
\left[
\log \frac{\mu}{\rho}
+
\sum_{n \not= 0}
e^{i n \frac{\tilde{x}_9}{\tilde{R}_9}}
K_0
\left(
\rho
\left|
\frac{n}{\tilde{R}_9}
\right|
\right)
\right].
\label{eq:codim2_reg}
\end{align}
As one sees, the $n \not= 0$ terms break isometry along the $\tilde{x}_9$-direction.
The two of the present authors studied the worldsheet instanton effects to the
$5^2_2$-brane geometry \cite{Kimura:2013zva}.
We have shown that the worldsheet instanton corrections to the exotic
$5^2_2$-brane geometry recover the $n\not=0$ isometry-breaking
contributions in \eqref{eq:codim2_reg}.
Therefore it is natural to expect that the corrections to the
logarithmic harmonic function in \eqref{eq:codim3_3} come from worldsheet
instantons stemming from F-strings wrapping on
two cycles associated with a $T^2$-fibration.
Indeed, the large-$\rho$ expansion of the
modified Bessel function shows the following relation:
\begin{align}
e^{
i m \frac{\tilde{x}_8}{\tilde{R}_8} + i n
 \frac{\tilde{x}_9}{\tilde{R}_9}
}
K_0
\left(
\rho
\sqrt{
\left(\frac{m}{\tilde{R}_8}\right)^2 +
\left(\frac{n}{\tilde{R}_9}\right)^2
}
\right)
\sim
\exp
\left[
- \rho
\sqrt{
\left(\frac{m}{\tilde{R}_8}\right)^2 +
\left(\frac{n}{\tilde{R}_9}\right)^2
}
+
i
\left(
m \frac{\tilde{x}_8}{\tilde{R}_8}
+
n \frac{\tilde{x}_9}{\tilde{R}_9}
\right)
\right].
\label{eq:exp_expression}
\end{align}
Here only the relevant terms are extracted.
The characteristic exponential behavior suggests that the winding
corrections in \eqref{eq:codim3_3} are interpreted as the instanton effects.
We will discuss this issue later.

\subsection{$R$-brane}
We proceed the T-duality chain.
We consider the periodic $5^2_2$-brane along the $x^7$-direction.
We begin with the same function as \eqref{eq:codim3_1} while the
variables $x^{8}$ and $R_{8}$ are replaced by $\tilde{x}_{8}$ and
$\tilde{R}_8$, respectively.
In order to introduce another extra isometry, we compactify the $x^7$-direction
to a circle $S^1$ of radius $R_7$ and consider a periodic array of the winding corrected $5^2_2$-brane.
The harmonic function becomes
\begin{align}
& H (\rho', x^7, \tilde{x}_8, \tilde{x}_9) = c +
\sum_{l=-\infty}^{\infty}
\sum_{m=-\infty}^{\infty}
\sum_{n=-\infty}^{\infty}
\frac{Q}{
\rho^{\prime 2}
+ (x^7 - 2 \pi R_7 l)^2 + (\tilde{x}_8 - 2 \pi \tilde{R}_8 m)^2 + (\tilde{x}_9 - 2 \pi \tilde{R}_9 n)^2}.
\label{eq:codim2}
\end{align}
Here $\rho' = x^6$.
We emphasize that the variables $\tilde{x}_{8,9}$ represent the
``winding'' coordinates in the $5^2_2$-brane configuration.
Once again by using the Poisson resummation \eqref{eq:Poisson}, we find
\begin{align}
H (\rho', x^7, \tilde{x}_8, \tilde{x}_9) =& \ c +
\frac{Q}{4 \pi R_7 \tilde{R}_8 \tilde{R}_9}
\sum_{l=-\infty}^{\infty}
\sum_{m=-\infty}^{\infty}
\sum_{n=-\infty}^{\infty}
e^{i l \frac{x^7}{R_7}}
e^{i m \frac{\tilde{x}_8}{\tilde{R}_8}}
e^{i n \frac{\tilde{x}_9}{\tilde{R}_9}}
\times
\notag \\
& \qquad \qquad
\times
\frac{1}{
\sqrt{
\left(
\frac{l}{R_7}
\right)^2
+
\left(
\frac{m}{\tilde{R}_8}
\right)^2
+
\left(
\frac{n}{\tilde{R}_9}
\right)^2
}
}
e^{- |\rho'|
\sqrt{
\left(
\frac{l}{R_7}
\right)^2
+
\left(
\frac{m}{\tilde{R}_8}
\right)^2
+
\left(
\frac{n}{\tilde{R}_9}
\right)^2
}
}.
\end{align}
Separating out the $l = m = n = 0$ sector, we have
\begin{align}
&
\lim_{l,m,n \to 0}
\frac{1}{
\sqrt{
\left(
\frac{l}{R_7}
\right)^2
+
\left(
\frac{m}{\tilde{R}_8}
\right)^2
+
\left(
\frac{n}{\tilde{R}_9}
\right)^2
}
}
e^{- |\rho'|
\sqrt{
\left(
\frac{l}{R_7}
\right)^2
+
\left(
\frac{m}{\tilde{R}_8}
\right)^2
+
\left(
\frac{n}{\tilde{R}_9}
\right)^2
}
}
\notag \\
\sim  &
- |\rho'|
\
+
\lim_{l,m,n \to 0}
\frac{1}{
\sqrt{
\left(
\frac{l}{R_7}
\right)^2
+
\left(
\frac{m}{\tilde{R}_8}
\right)^2
+
\left(
\frac{n}{\tilde{R}_9}
\right)^2
}
}
+ \cdots,
\end{align}
where $\cdots$ vanishes in the limit $l,m,n \to 0$.
Again, the divergent part comes from this sector.
Any regulator that behaves linearly divergent works well.
Subtracting this term 
in \eqref{eq:codim2} 
and performing the generalized T-duality
transformation by the replacement $x^7 \to \tilde{x}_7$ and $R_7 \to \tilde{R}_7$,
we find the harmonic function of the $5^3_2$-brane
\begin{align}
H (\rho', \tilde{x}_7, \tilde{x}_8, \tilde{x}_9) =& \
\text{const.}
-
\frac{Q}{4\pi \tilde{R}_7 \tilde{R}_8 \tilde{R}_9}
|\rho'|
+
\frac{Q}{4 \pi \tilde{R}_7 \tilde{R}_8 \tilde{R}_9}
\sum_{l,m,n \not= (0,0,0)}
e^{i l \frac{\tilde{x}_7}{\tilde{R}_7}}
e^{i m \frac{\tilde{x}_8}{\tilde{R}_8}}
e^{i n \frac{\tilde{x}_9}{\tilde{R}_9}}
\times
\notag \\
&
\times
\frac{1}{
\sqrt{
\left(
\frac{l}{\tilde{R}_7}
\right)^2
+
\left(
\frac{m}{\tilde{R}_8}
\right)^2
+
\left(
\frac{n}{\tilde{R}_9}
\right)^2
}
}
e^{- 
|\rho'| 
\sqrt{
\left(
\frac{l}{\tilde{R}_7}
\right)^2
+
\left(
\frac{m}{\tilde{R}_8}
\right)^2
+
\left(
\frac{n}{\tilde{R}_9}
\right)^2
}
}.
\label{eq:codim2_1}
\end{align}
In the limit $\tilde{R}_{7,8,9} \to 0$,
the linear term in $\rho' = x^6$ is the harmonic function for
branes of codimension one ({\it i.e.} domain walls)
\begin{align}
H (x^6,\tilde{x}_7, \tilde{x}_8, \tilde{x}_9) \to \text{const.}
- \frac{Q}{4 \pi \tilde{R}_7 \tilde{R}_8 \tilde{R}_9} |x^6|.
\end{align}
The $l,n,m \not=0$ parts are winding corrections to the domain wall solution.
Again we expect that this comes from the instanton stemming from the
F-strings wrapping on three circles associated with a $T^3$-fibration.
This harmonic function appeared in the DFT solution 
(see the footnote \ref{ft:cd1})
and reproduces the one discussed in \cite{Lust:2017jox}.

\subsection{Space-filling brane in winding space}
We finally compactify the remaining transverse direction of the
$5^3_2$-brane background.
We begin with the same function as \eqref{eq:codim2} where the variables $x^7$ and $R_7$ are replaced by $\tilde{x}_7$ and $\tilde{R}_7$.
We compactify the $x^6$-direction
to a circle $S^1$ of radius $R_6$ and consider a periodic array of the winding corrected $5^3_2$-brane.
The harmonic function becomes
\begin{align}
H(x^6,\tilde{x}_7, \tilde{x}_8, \tilde{x}_9) =& \ c +
\sum_{k=-\infty}^{\infty}
\sum_{l=-\infty}^{\infty}
\sum_{m=-\infty}^{\infty}
\sum_{n=-\infty}^{\infty}
\times
\notag \\
&
\times
\frac{Q}{(x^6 - 2 \pi R_6 k)^2 + (\tilde{x}_7 - 2 \pi \tilde{R}_7 l)^2 +
 (\tilde{x}_8 - 2 \pi \tilde{R}_8 m)^2 + (\tilde{x}_9 - 2 \pi
 \tilde{R}_9 n)^2}.
\end{align}
Then, after the Poisson resummation \eqref{eq:Poisson}, we have
\begin{align}
H(x^6,\tilde{x}_7, \tilde{x}_8, \tilde{x}_9) =& \
 c +
\frac{Q}{4 \pi^2 R_6 \tilde{R}_7 \tilde{R}_8 \tilde{R}_9}
\sum_{k=-\infty}^{\infty}
\sum_{l=-\infty}^{\infty}
\sum_{m=-\infty}^{\infty}
\sum_{n=-\infty}^{\infty} \times
\notag \\
&
\times
e^{i k \frac{x^6}{R_6}}
e^{i l \frac{\tilde{x}_7}{\tilde{R}_7}}
e^{i m \frac{\tilde{x}_8}{\tilde{R}_8}}
e^{i n \frac{\tilde{x}_9}{\tilde{R}_9}}
\left[
\left(
\frac{k}{R_6}
\right)^2
+
\left(
\frac{l}{\tilde{R}_7}
\right)^2
+
\left(
\frac{m}{\tilde{R}_8}
\right)^2
+
\left(
\frac{n}{\tilde{R}_9}
\right)^2
\right]^{-1}.
\end{align}
A regulator that diverges quadratically works sufficiently. 
Subtracting the divergent piece in the $k=l=m=n=0$ sector, and
performing the formal T-duality transformation by $x^6 \to \tilde{x}_6$ and $R_6 \to \tilde{R}_6$,
we have
\begin{align}
H(\tilde{x}_6,\tilde{x}_7, \tilde{x}_8, \tilde{x}_9) =& \
\text{const.}
+
\frac{Q}{4 \pi^2 \tilde{R}_6 \tilde{R}_7 \tilde{R}_8 \tilde{R}_9}
\sum_{k,l,m,n \not= (0,0,0,0)}
e^{i k \frac{\tilde{x}_6}{\tilde{R}_6}}
e^{i l \frac{\tilde{x}_7}{\tilde{R}_7}}
e^{i m \frac{\tilde{x}_8}{\tilde{R}_8}}
e^{i n \frac{\tilde{x}_9}{\tilde{R}_9}}
\times
\notag \\
&
\times
\left[
\left(
\frac{k}{\tilde{R}_6}
\right)^2
+
\left(
\frac{l}{\tilde{R}_7}
\right)^2
+
\left(
\frac{m}{\tilde{R}_8}
\right)^2
+
\left(
\frac{n}{\tilde{R}_9}
\right)^2
\right]^{-1}.
\label{eq:space-filling}
\end{align}
The constant term is the harmonic function for
space-filling branes of codimension zero.
This is a source of the mixed symmetric potential $D_{abcd}$ \cite{Bergshoeff:2016ncb},
in which a fully localized space-filling brane in six dimensions is predicted.
Conventional geometry is realized as the ``zero-winding'' sector.
The winding corrections would be the F-string wrapping on $T^4$.
Indeed, if we consider the smearing limit $\tilde{R}_{6,7,8,9} \to 0$,
then
\begin{align}
H(\tilde{x}_6,\tilde{x}_7, \tilde{x}_8, \tilde{x}_9) \to \text{const}.
\end{align}

Our analysis suggests that even though space-filling branes have trivial
space-time dependence, it can have non-trivial winding structures due
to stringy effects, even though space-filling branes are consistent solutions only when
certain kind of a sink of brane charges (like the orientifold plane
which has the negative D-brane charge) are simultaneously placed.
One finds that there are no ``damping factors'' in the exponential
part in \eqref{eq:space-filling}.
This is contrasted with the higher codimensional cases.
The reason behind this will be explained later.

\subsection{Wave and F-string}
We now turn to the DFT wave solutions.
The discussion is parallel to the five-brane case.
The harmonic function of the F-string compactified in the $x^8$-direction is given by
\begin{align}
H(x^8, r)
= c +
\sum_{n=-\infty}^{\infty}
\frac{h}{(r^2 + (x^8 + 2\pi n R_8)^2)^3}
\end{align}
where
$r^2 = \delta_{i'j'} x^{i'} x^{j'}$ ($i' = 1,\ldots,7$)
and $R_8$ is the radius in the $x^8$-direction.
Rewriting the discrete modes, and performing the formal T-duality
transformation, we find
\begin{align}
H(r,\tilde{x}_8) = c + \frac{3h}{16r^5} \sum_{n=-\infty}^\infty
	\left[
	1 + \frac{n}{\tilde{R}_8} r + \frac{n^2}{\tilde{R}_8^2} \frac{r^2}{3}
	\right]
	e^{
i n \frac{\tilde{x}_8}{\tilde{R}_8} } e^{-r \left|
 \frac{n}{\tilde{R}_8} \right|.}
\label{eq:F1w1}
\end{align}
Here $\tilde{x}_8$ is the winding coordinate.
We notice that the characteristic exponential structure again appears
as the winding corrections.
This fact is an indication of the string worldsheet instanton corrections
to the geometry.

Further compactifications are also possible but
calculations of the Poisson resummation in lower dimensions are rather complicated.
Instead, we switch to a direct manipulation of the harmonic function and
discuss the general property of it.
Let us consider the $(d+N)$-dimensional transverse space of an extended object (branes,
F-strings and so on).
We then assume that a part of the space is compactified on the $N$-dimensional torus
$T^N$ and the extended object is T-dualized along
that directions.
The Laplacian in the transverse space is decomposed as
\begin{align}
\Box_{d + N} = \Box_d + \TildeBox_N,
\end{align}
where $\Box_d$ is the ordinary Laplacian defined by the $d$-dimensional
geometric coordinates $x^m \ (m=1, \ldots, d)$ while $\TildeBox_N$ is defined by the
winding coordinates $\tilde{x}_i \ (i= 1, \ldots, N)$.
Employing the spherically symmetric configuration in the geometric
sector, the harmonic function $H$ satisfies
\begin{align}
\Box_{d + N} H (r,\tilde{x}) = 0, \quad r^2 = (x^1)^2 + \cdots
+ (x^d)^2.
\label{eq:laplace_eq}
\end{align}
We look for the solution to this equation. Considering the ansatz $H
(r,\tilde{x}) = f(r) \varphi (\tilde{x})$, the equation
\eqref{eq:laplace_eq} reduces to
\begin{align}
\frac{1}{r^{d-1} f} \frac{\partial}{\partial r}
\left(
r^{d-1} \frac{\partial}{\partial r}
\right) f (r) +
\frac{1}{\varphi (\tilde{x})} \TildeBox_N \varphi
 (\tilde{x}) = 0.
\end{align}
This equation is solved by
\begin{align}
&
\TildeBox_N \varphi (\tilde{x}) = - \lambda^2 \varphi (\tilde{x}),
\label{eq:laplace1}
\\
&
r^{d-1} f''(r) + (d-1) r^{d-2} f' (r) - \lambda^2 r^{d-1} f (r) = 0,
\label{eq:laplace2}
\end{align}
where $\lambda$ is a constant.
Since the $\tilde{x}_i$-directions are compactified, $\varphi$
should satisfy the periodic condition
\begin{align}
\varphi (\tilde{x}_i + 2 \pi R_i n_i) = \varphi (\tilde{x}_i),
\quad n_i \in \mathbb{Z},
\label{eq:periodic}
\end{align}
where we have assumed that the radius of each circle in $T^N$ is $R_i$.
A function that satisfies the equation \eqref{eq:laplace1} and the
condition \eqref{eq:periodic} is easily found to be
\begin{align}
\varphi_{\vec{n}} = C \exp
\left[
- i \sum_{i=1}^N n_i \frac{\tilde{x}_i}{R_i}
\right], \quad
\lambda_{\vec{n}}^2 =
\sum_{i=1}^N
\left(
\frac{n_i}{R_i}
\right)^2,
\end{align}
where $C$ is a constant.
It is straightforward to show that the solution to the equation
\eqref{eq:laplace2} is given by
\begin{align}
f (r) = c_1 \, r^{\frac{2-d}{2}} \, J_{(d-2)/2} (- i r \lambda_{\vec{n}})
+ c_2 \, r^{\frac{2-d}{2}} \, Y_{(d-2)/2} (- i r \lambda_{\vec{n}}),
\end{align}
where $c_{1,2}$ are constants and $J_{\nu} (x), Y_{\nu} (x)$ are Bessel functions of the first and
the second kinds.
The general solution is given by $H (r,\tilde{x}) = \sum_{\vec{n}} f (r)
\varphi_{\vec{n}} (\tilde{x})$.
One finds that the large-$r$ expansion of the Bessel functions again
exhibits the characteristic exponential behavior appeared in \eqref{eq:exp_expression}.

\section{Worldsheet instanton effects} \label{sect:instanton}
In this section, we discuss an interpretation of the winding corrections
to geometries as the string worldsheet instanton effects.
The worldsheet instantons are configurations that minimize
the Euclidean action of the fundamental string in a given topological
sector \cite{Wen:1985jz}.
This is a map $\phi$ from the tree-level worldsheet $\Sigma = S^2$ to a
two-cycle $K$ in the target space which minimizes the
worldsheet Polyakov action:
\begin{align}
S = \frac{1}{4 \pi \alpha'} \int_{\Sigma} d^2 \sigma \, \sqrt{h}
\, g_{\mu \nu} \, \frac{\partial X^{\mu}}{\partial \sigma^a} \frac{\partial
 X^{\nu}}{\partial \sigma^b} \, h^{ab}.
\end{align}
Here $\sigma^a \ (a=1,2)$
are the coordinates of the Euclidean worldsheet
and $h_{ab}$ is the worldsheet metric,
whilst $X^{\mu} \ (\mu = 0, \ldots, 9)$ are the worldsheet scalars
and $g_{\mu \nu}$ is the metric of the target space-time.
One finds that the following inequality holds:
\begin{align}
S \ge
\mp \frac{1}{4 \pi \alpha'} \int_{\Sigma} \! d^2 \sigma \,
\varepsilon^{ab} \, \Omega_{\mu \nu} \, \partial_a X^{\mu} \partial_b X^{\nu},
\label{eq:topological_inv}
\end{align}
where $\varepsilon^{ab}$ is the Levi-Civita symbol on the worldsheet
while $J^{\mu}{}_{\nu}$ and $\Omega_{\mu \nu} = g_{\mu \rho}
J^{\rho}{}_{\nu}$ are a complex structure and a K\"{a}hler form in the
target space, respectively.
The equality is saturated when the following condition is satisfied:
\begin{align}
\partial_a X^{\mu} \pm J^{\mu} {}_{\nu} \, \varepsilon_{ab} \, \partial^b X^{\nu} = 0.
\label{eq:NLSM_inst}
\end{align}
This is the instanton equation and its solution is given by a
(anti) holomorphic map.
The worldsheet Polyakov action for the solution to \eqref{eq:NLSM_inst}
is evaluated as
\begin{align}
S =
\mp \frac{1}{4 \pi \alpha'} \int_{\Sigma} d^2 \sigma \,
 \varepsilon^{ab} \, \Omega_{\mu \nu} \, \partial_a X^{\mu} \partial_b X^{\nu}
=
\mp \frac{1}{4 \pi \alpha'} \int_{\Sigma} \phi^{*} (\Omega).
\label{eq:NLSM_bound}
\end{align}
Here $\phi^{*}$ is the pull-back from $K$ to
$\Sigma$.
The integration \eqref{eq:NLSM_bound} provides a topological invariant
associated with the homology group $H_2 (S^2) = \mathbb{Z}$.
This map is classified by the homotopy group $\pi_2 (K)$.

We note that even for geometries where no non-trivial
two-cycles exist, it is formally
possible to consider the worldsheet instantons \cite{Tong:2002rq}.
These can be interpreted as a ``point-like instanton''
\cite{Witten:1993yc,Witten:1995gx}. However, in some cases, one can define non-trivial
two-cycles in an appropriate limit of parameters that specify geometries.
In these cases, one defines the worldsheet instantons as a limit of
``disk instantons'' \cite{Okuyama:2005gx}.
In the following, we discuss these issues.

\subsection{Disk instantons in single-centered Taub-NUT space}
We now focus on the background geometry of the KK-monopole.
The metric is
\begin{align}
ds^2 = dx^2_{012345} + H(r) dx^2_{678} + H^{-1} (r) (dx^9 + A_i d x^i)^2,
 \quad i= 6,7,8,
\end{align}
where
$x^{0,1,2,3,4,5}$ are the world-volume directions
while the transverse space
given by $x^{6,7,8,9}$ is represented by the Taub-NUT space.
The $x^9$-direction is compactified on $S^1$ and the $U(1)$ isometry is
defined on it.
$A_i$ is the KK gauge field.
The harmonic function $H(r)$ is given by
\begin{align}
H (r) = \frac{1}{g^2} + \sum_{i} \frac{Q}{r_i}, \quad r^2_i = (x^6 -
 x^6_i)^2 + (x^7 - x^7_i)^2 + (x^8 - x^8_i)^2.
\end{align}
Here the constant $g$ is the string sigma model coupling
which determines the
asymptotic radius of the $S^1$-fiber.
The constants $x_i^{6,7,8}$ specify the centers of the Taub-NUT space.
The $S^1$ in the $x^9$-direction is fibered on the base space
spanned by $x^{6,7,8}$.
More precisely, an $r = \text{const.}$ surface in $x^{6,7,8}$ space
represents a two-sphere $S^2$ with radius $r$ and $S^1$ is fibered on
there making topologically $S^3$. This is the Hopf fibration $S^1$ over
$S^2$.

The physical radius of the fibered $S^1$ is given by $H^{-1}$.
In the multi-centered Taub-NUT space, at each center
$\vec{r}_i = (x^6 - x^6_i, x^7 - x^7_i, x^8 - x^8_i) = 0$
we find the fibered $S^1$ shrinks to zero size.
We therefore find a two-cycle $S_{ij}$
defined on a line segment between two centers $\vec{r}_{ij} = \vec{r}_j - \vec{r}_i$
whose topology is two-sphere $S^2$, see Figure \ref{fig:2-cycle}.
\begin{figure}[tb]
\begin{center}
\includegraphics[scale=1.5]{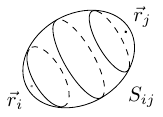}
\end{center}
\caption{A two-cycle defined on the multi-centered Taub-NUT space.}
\label{fig:2-cycle}
\end{figure}
In this case, the worldsheet instanton is defined
as the mapping from $\Sigma = S^2$ to $S_{ij}$, where
the instanton number is assigned by the homotopy $\pi_2 (S^2) = \mathbb{Z}$.

On the other hand, for the single centered Taub-NUT space,
we have a vanishing $S^1$ at the center $r=0$ and the radius of $S^1$ at
$r \to \infty$ is finite,
\begin{align}
H^{-1} (0) = 0, \quad H^{-1} (\infty) = g^2.
\end{align}
Now, the $S^1$-fiber over the segment between $r = 0$ and $r = \infty$
becomes an open cigar whose topology is disk $D^2$.
Therefore there are no two-cycles that define the worldsheet instantons
in the single-centered Taub-NUT space.
However, if we recognize that the worldsheet $\Sigma = S^2$ as the set of disk $D^2$ and infinity:
\begin{align}
\Sigma = D^2_{\Sigma} \cup \{\infty_{\Sigma} \},
\end{align}
we can define a map from $D^2_{\Sigma}$ to $D^2$,
while $\{\infty_{\Sigma}\}$ to space-time
infinity $\{\infty \}$ as \cite{Okuyama:2005gx}:
\begin{align}
D^2_{\Sigma} = \{z \in \mathbb{C} : |z| < 1 \} \
\longrightarrow \
n C_i = \{ \vec{r} (z) = \vec{r}_i + f (|z|) \vec{v},
\ x^9(z) = n  \arg (z)\},
\label{eq:map_disk}
\end{align}
where $C_i$ is a cigar whose tip is given by $\vec{r}_i$,
, the unit vector $\vec{v}$ defines a direction of the segment and
$x^9$ is the coordinate of the fibered $S^1$.
The integer $n$ is the winding number and $f(|z|)$ is a function which satisfies the boundary condition
\begin{align}
f(0) = 0, \quad f(1) = \infty.
\end{align}
We identify $z = 1$ with $\infty_{\Sigma}$.
When we go around the origin inside the disk at some fixed radius
 $|z| = A = \text{fixed} < 1$,
the map winds $S^1$ $n$-times. see Figure \ref{fig:disk_inst}.
This map is called the disk instanton.
\begin{figure}[tb]
\begin{center}
\includegraphics
{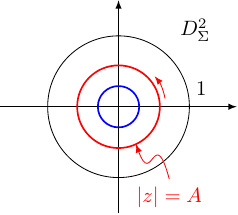}
\hspace{20pt}
\raisebox{17mm}{$\overset{\text{mapping}}{\Longrightarrow}$}
\hspace{20pt}
\raisebox{-5mm}{\includegraphics
{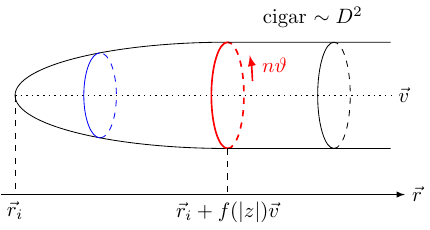}}
\end{center}
\caption{The map from worldsheet disk to the one in space-time.}
\label{fig:disk_inst}
\end{figure}

One should remember that the instanton calculus in the GLSM setup is performed
in the limit $g \to 0$ \cite{Harvey:2005ab}.
This limit implies that the cigar defined on the single-centered Taub-NUT
space is closed at infinity.
Therefore the cigar in this limit becomes topologically
$S^2$.
Schematically we have the following relation:
\begin{align}
[\text{Disk instanton}]
\quad
\xrightarrow[g \to 0]{}
\quad
[\text{Worldsheet instanton}].
\end{align}

\subsection{Worldsheet instanton corrections to the $5^2_2$-brane geometry}
Following the discussion in the previous subsection, we proceed to the
$5^2_2$-brane geometry \cite{deBoer:2010ud}:
\begin{align}
& ds^2 = dx^2_{012345} + H dx^2_{67} + HK^{-1}
((dx^8)^2 + (dx^9)^2),
\notag \\
& H = h_0 + \sigma \log \frac{\mu}{\rho},
\quad \rho^2 =
(x^6)^2 + (x^7)^2,
\notag \\
& B_{89} = K^{-1} A_8, \quad K = H^2 + A_8^2, \quad e^{2\phi} = H
 K^{-1},
\notag \\
& A_8 = c \ \text{arctan}
\left(
\frac{x^7}{x^6}
\right) + \text{const.},
\end{align}
where $h_0, \sigma, \mu, c$ are constants.
Now there are two isometries defined in the $x^{8,9}$-directions.
The situation is similar to the
one in the Taub-NUT space, {\it i.e.} the two-torus
$T^2 = S^1 \times S^1$ is fibered over the two-dimensional
space $\mathbb{R}^2$.
The physical radius of each $S^1$ is given by $HK^{-1}$.
At the origin $\rho \to 0$ in the $x^{6,7}$-plane,
we have $HK^{-1} \to 0$, while at the cutoff scale
$\rho = \mu$, $HK^{-1} = \frac{h_0}{h_0^2 + A_8^2}< \infty$
is finite provided the constant $h_0$ is finite.
Therefore, the volume of the fibered torus vanishes at the origin.
We find that the two circles $S_i^1 \ (i=1,2)$ are
separately fibered over a segment defined between $\rho = \rho_0 = 0$ and the cutoff surface
$\rho =\mu$ in the $x^{6,7}$-plane.
Here $i = 1, 2$ labels the two circle directions $x^8,x^9$.

Since the one-dimensional homology group of the two torus is
$H_1 (T^2) = \mathbb{Z} \oplus \mathbb{Z}$,
a general one-cycle in $T^2$ is generated by two integers.
There are two independent one-cycles in each $S^1$
in $T^2 = S^1 \times S^1$.
One can then define a one-cycle by a formal linear combination
of these two one-cycles.
For example, a one-dimensional string that wraps on the first $S^1$
$n_1$-times and the second $S^1$ $n_2$-times defines a one-cycle in
$T^2$. We call this the $(n_1,n_2)$-cycle which is completely classified by two integers.
Analogous to the Taub-NUT case,
this $(n_1,n_2)$-cycle is fibered over the segment $\rho \in [0, \mu]$
and defines an open cigar.
This naturally induces a map from the worldsheet disk $D^2_{\Sigma}$ to
the disk $D^2$ defined over the $(n_1,n_2)$-cycle fibration:
\begin{align}
&
D^2_{\Sigma} =
\left\{
z \in \mathbb{C} : |z| < 1
\right\}
\notag \\
& \longrightarrow
(n_1,n_2) C =
\left\{
\vec{\rho} (z) = \vec{\rho}_0 + f(|z|) \vec{v},
\
x^8 (z) = n_1 \text{arg} (z),
x^9 (z) = n_2 \text{arg} (z)
\right\},
\notag \\
&
f(0) = 0, \quad f(1) = \mu.
\end{align}
Namely, when one goes around the origin inside the worldsheet disk, the
map winds the first $S^1$ $n_1$-times and the second $S^1$ $n_2$-times.
This fact substantially leads to a generalization of the disk instantons.
See Figure \ref{fig:522_cigars}.
\begin{figure}[tb]
\begin{center}
\includegraphics
{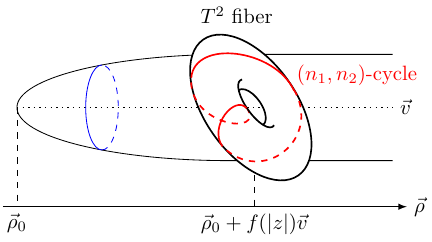}
\end{center}
\caption{Open cigar in the $5^2_2$-brane geometry
whose boundary is given by the $(n_1, n_2)$-cycle.
}
\label{fig:522_cigars}
\end{figure}
If we consider the limit $h_0 \to \infty$,
then the two open cigars $D^2_i$ are closed at $\rho \sim \mu$
forming $S^2_i$.
It is therefore obvious that the map is classified by the homotopy group
$\pi_2 (S^2_{1}) \oplus \pi_2 (S^2_{2}) = \mathbb{Z} \oplus \mathbb{Z}$.

In order to justify the above homotopy assignment and
define the topological charges, we examine the homology
structure of the two-cycles.
Since we have the Betti number
$b_2 (S^2) = 1$, we can define a harmonic $(1,1)$-form
$b^i (i=1,2)$ in each $S^2_i$ for which we have the period matrix:
\begin{align}
\int_{S^2_i} b^j = \tilde{R}_i^{-1} \delta_i {}^j,
\quad (\text{$i$: no sum}).
\label{eq:period_matrix}
\end{align}
Here $\tilde{R}_i$ are constants.
Then the K\"{a}hler form $\Omega$ is expanded by
\begin{align}
\Omega = \sum_{i=1,2} c_i b^{i},
\end{align}
where $c_i$ are real constants.
On the other hand, corresponding to the $(n_1,n_2)$-cycle, we can associate a linear
combination of closed cigar $S^2 = n_1 S^2_1 + n_2 S^2_2$.
Utilizing these facts, the BPS bound of the worldsheet Polyakov action
is given by
\begin{align}
S \ge \frac{1}{4 \pi \alpha'} \vec{c} \cdot \vec{Q},
\end{align}
where $\vec{c} = (c_1,c_2)$, $\vec{Q} = (Q^1, Q^2)$ and
\begin{align}
 Q^i = \int_{\Sigma} \phi^* (b^i) =  \int_{\phi (\Sigma)} b^i = \frac{n_i}{\tilde{R}_i},
\end{align}
where we have used $\phi (\Sigma) = n_1  S^2_1 + n_2 S^2_2$ and the
relation \eqref{eq:period_matrix}.
This is the topological charge \cite{Ruback:1987sg, Dine:1987bq}
associated with the map defined above.
The most stringent bound of the action is found to be
\begin{align}
 S \ge \frac{1}{4 \pi \alpha'} |\vec{c}|
\sqrt{
\left(\frac{n_1}{\tilde{R}_1} \right)^2 + \left(\frac{n_2}{\tilde{R}_2} \right)^2
}.
\end{align}
This bound is saturated when the condition \eqref{eq:NLSM_inst} is
satisfied and the vector $\vec{c}$ is parallel to $\vec{Q}$.
Introducing the contribution of the Kalb-Ramond field,
we have the instanton action (where we have chosen the plus sign in
\eqref{eq:NLSM_bound}):
\begin{align}
S_{\text{inst.}} =
\frac{1}{4 \pi \alpha'}
\left[
|\vec{c}|
\sqrt{
\left(\frac{n_1}{\tilde{R}_1} \right)^2 + \left(\frac{n_2}{\tilde{R}_2} \right)^2
}
- i \vec{n} \cdot \vec{\mathcal{B}}
\right].
\end{align}
Here
\begin{align}
\mathcal{B}^i = - \int_{\phi^{*} (S^2_i)} \! d^2 \sigma \, B_{\mu} {}^{\rho}
\, \Omega_{\rho \nu} \, \partial^a X^{\mu} \partial_a X^{\nu}.
\end{align}
Assigning $|\vec{c}|$ the distance scale $\rho$, we find the precise
agreement between $e^{-S_{\text{inst.}}}$ and the exponential factor
appeared in \eqref{eq:exp_expression}.
This is a conceivable result.

One then proceeds the cases with more isometries, for example,
the $5^3_2$-branes ($R$-brane). The discussion is completely parallel.
For example, the geometry of the $5^3_2$-branes includes $T^3$-fibration
over $\mathbb{R}$. The homology $H_1 (T^3) = \mathbb{Z} \oplus
\mathbb{Z} \oplus \mathbb{Z}$ indicates that disk instantons are
labelled by three integers $(n_1,n_2,n_3)$. Indeed, we can define a
generalization of the map \eqref{eq:map_disk} and we deduce that
instanton action agrees with the exponential factor appeared in
\eqref{eq:codim2_1}.
The same is true for the DFT wave solutions.
However, this picture fails when we move to space-filling branes.
There are no non-trivial base space in this case.
Therefore, there is no straightforward way to generalize the discussion
to space-filling branes.

\section{Instantons in GLSM} \label{sect:GLSM}
In this section, we consider
the worldsheet instantons in the GLSM language.
The explicit instanton calculus was first performed in
\cite{Tong:2002rq} in which, in order to realize the H-monopole geometry,
the two-dimensional $\mathcal{N} = (4,4)$ vector, hyper and twisted
hypermultiplets are prepared.
The gauge group is $U(1)$ which has a direct connection to the isometry
of the H-monopole.
As we have mentioned before, the instanton calculus is justified only when the sigma model coupling
is small $g \to 0$. In this limit, the GLSM is reduced to a truncated model. For the GLSMs describing the H-monopole and KK-monopole, the
truncated model is given by the Abelian-Higgs model.
Instantons in the GLSM is the BPS vortices associated with the $U(1)$ gauge field \cite{Tong:2002rq}.
This is indeed the case for the $5^2_2$-brane geometry
\cite{Kimura:2013fda, Kimura:2013zva}\footnote{However, we have realized only the one isometry of the $5^2_2$-brane in \cite{Kimura:2013fda}. A GLSM that has $U(1)^2$ isometry is studied in \cite{Kimura:2013khz}.}.

Now we consider GLSM describing geometries with $T^N$-fibration.
Since the isometry of the background geometry is encoded into the gauge
symmetry in the GLSM, we assume that a conceivable gauge group is
$G = U(1) \times U(1) \times \cdots = U(1)^N$ which is associated with the isometry in $T^N$.
Following the previous works for the H-, KK-monopoles and $5^2_2$-brane, we also assume
that the truncated model in an appropriate limit of parameters
is given by the $(2+0)$-dimensional Abelian-Higgs model,
\begin{align}
\mathcal{L}_{\text{E}}
= \sum_{a=1}^N
\left[
\frac{1}{2 e_a^2} (F_{12,a})^2 + |D_m q_a|^2 + \frac{e_a^2}{2}
\left(
|q_a|^2 - \sqrt{2} \, \zeta_a
\right)^2 + i \sqrt{2} \, \vartheta_a F_{12,a}
\right].
\label{eq:truncated_model}
\end{align}
Here $A_{m,a} \ (m=1,2; a=1, \ldots,N)$ are the $U(1)$ gauge fields,
$q_a$ are charged scalar fields, $\zeta_a$ are the Fayet-Iliopoulos (FI)
parameters associated with each $U(1)$ sector.
The covariant derivatives and the gauge field strength are given as
$D_m q_a = \partial_m q_a + i A_{m,a} q_a \ (m=1,2; a:\text{no sum})$ and
$F_{12,a} = \partial_1 A_{2,a} - \partial_2 A_{1,a}$, respectively.
The last term is a topological term for which $\vartheta_a \ (a=1,2,\ldots, N)$ is scalar fields corresponding to the winding
coordinates.
For the KK-monopole and $5^2_2$-brane cases, the charged scalar field
comes from the hypermultiplet while the scalar field $\vartheta$ is a
``remnant'' of the twisted hypermultiplet presented in the GLSM for the H-monopole.
It has been shown that the topological term including $\vartheta$ is left behind in the
process of the T-duality transformation \cite{Rocek:1991ps}.
In the T-dualized frame, this $\vartheta$ can be interpreted as the dual
coordinate. In other words, this is the GLSM origin of the winding coordinate dependence
of geometries.

It is easy to find the Bogomol'nyi bound of the Euclidean Lagrangian
\eqref{eq:truncated_model}. We find
\begin{align}
\mathcal{L}_{\text{E}} \ge& \
\mp \sqrt{2} \sum_a \zeta_a F_{12,a} + i \sqrt{2} \sum_a \vartheta_a
 F_{12,a}
+ \text{(total derivative)}.
\end{align}
If we define $\vec{\zeta} = (\zeta_{a=1}, \zeta_{a=2}, \ldots)$,
$\vec{\vartheta} = (\vartheta_{a=1}, \vartheta_{a=2}, \ldots)$,
$\vec{F}_{12} = (F_{12,a=1}, F_{12,a=2}, \ldots)$,
and
$\vec{Q} \equiv - \frac{1}{2\pi} \int \! d^2 x \ \vec{F}_{12}$,
the most stringent bound of the Euclidean action
$S_{\text{E}} = \frac{1}{2\pi} \int \! d^2 x \ \mathcal{L}_{\text{E}}$ is
\begin{align}
S_{\text{E}} \ge & \  \sqrt{2} |\vec{\zeta}|
\sqrt{
\sum_a Q_a^2
}
- i \sqrt{2} \, \vec{\vartheta} \cdot \vec{Q}.
\end{align}
Here the equality is saturated when the FI parameter $\vec{\zeta}$ and
$\vec{Q}$ becomes parallel $\vec{\zeta} \parallel \vec{Q}$ and
the following BPS equations hold:
\begin{align}
F_{12,a} \mp e_a^2 (|q_a|^2 - \sqrt{2} \, \zeta_a) = 0,
\quad
D_1 q_a \pm i D_2 q_a = 0,
\qquad a = 1,2, \ldots, N.
\label{eq:ANO_vortex}
\end{align}
The equations \eqref{eq:ANO_vortex} are nothing but the
Abrikosov-Nielsen-Olesen (ANO) vortex equations in each $U(1)$ sector ($a=1,2, \ldots, N$).
They are independent equations and give the non-trivial
first Chern numbers specified by the homotopy group
$\pi_1 (G) = \pi_1 (S^1) \oplus \pi_1 (S^1) \oplus \cdots \simeq
\mathbb{Z}^N$.
The topological charges are given by
\begin{align}
\vec{Q} = (n_1, n_2, \ldots, n_N),
\quad
n_a \in \mathbb{Z},
\qquad a =1, \ldots, N.
\end{align}
The path integral on the solution to \eqref{eq:ANO_vortex} is calculated
by the integral over the moduli space
$\mathcal{M}_{n_1, n_2, \ldots, n_N}$
of each ANO vortex.
Rescaling the gauge field $A_{m,a} \to \frac{1}{\tilde{R}_a}
A_{m,a}$ and taking into account the overall factor and rewrite
$\vartheta_a \equiv \tilde{x}_a$, $|\vec{\zeta}| \equiv \rho$
\cite{Tong:2002rq}, we obtain (we have chosen the plus sign in \eqref{eq:ANO_vortex})
\begin{align}
Z
= \sum_{n_1, n_2, \ldots, n_N} \int
 \! d \mathcal{M}_{n_1, \ldots, n_N} \exp
\left[
- \rho
\sqrt{
\left(\frac{n_1}{\tilde{R}_1}\right)^2
+
\left(\frac{n_2}{\tilde{R}_2}\right)^2
+ \cdots
}
+ i
\left(
n_1 \frac{\tilde{x}_1}{\tilde{R}_1}
+
n_2 \frac{\tilde{x}_2}{\tilde{R}_2}
+ \cdots
\right)
+ \cdots
\right],
\label{eq:inst_p.f.}
\end{align}
where the last $\cdots$ in the exponential factor stands for the moduli
of the vortices.
This again produces the characteristic structure appeared in
\eqref{eq:exp_expression}. Therefore the winding corrections found in
the DFT solutions are expected to be realized as vortex corrections in
GLSM. The related group structures of the instantons are found in Table \ref{tb:NLSM_GLSM}.

\begin{table}[tb]
\centering
  \begin{tabular}{c||c|c}
 & KK-monopole & $5^2_2$-brane  \\
\hline \hline
Fibration & $S^1$ over $\mathbb{R}^{3}$ & $T^2 = S^1 \times S^1$ over
	   $\mathbb{R}^{2}$ \\
\hline
Homology & $H_1 (S^1) = \mathbb{Z}$ & $H_1 (T^2) = \mathbb{Z}
	   \oplus \mathbb{Z}$ \\
\hline
NLSM homotopy & $\pi_2 (S^2) = \mathbb{Z}$ & $\pi_2 (S^2) \oplus \pi_2 (S^2) =
	   \mathbb{Z} \oplus \mathbb{Z}$  \\
\hline
GLSM homotopy & $\pi_1 (U(1)) = \mathbb{Z}$ & $\pi_1 (U(1)^2) =
	   \mathbb{Z} \oplus \mathbb{Z}$ \\
  \end{tabular}
\caption{
Related group structures of the instantons in NLSM and GLSM.
}
\label{tb:NLSM_GLSM}
\end{table}

Finally, we comment on the complete description of the Lagrangian.
In this paper we focused only on the topological structure of the target space, i.e., we neglected any contributions from F-terms which play a significant role in constructing supersymmetric theory and describing the target space geometry. Since the target space is a T-dualized version of the Taub-NUT space, the gauge theory must have ${\cal N}=(4,4)$ supersymmetry which requires $SU(2)$ R-symmetry. However, more than one topological terms such as $\sum_a \vartheta_a F_{12,a}$ would break the $SU(2)$ R-symmetry to $U(1)$, hence the ${\cal N}=(4,4)$ supersymmetry is broken to ${\cal N}=(2,2)$
at least in the classical level. The reason is as follows.
We begin with the GLSM for five-branes \cite{Tong:2002rq} (and series of
works \cite{Kimura:2013fda, Kimura:2013zva, Kimura:2013khz,
Kimura:2014aja, Kimura:2015yla, Kimura:2015cza, Kimura:2015qze}).
Four transverse directions of the five-branes are represented by four scalars belonging to the twisted hypermultiplet.
$\vartheta_a$ is chosen as one component of this multiplet.
The remaining three scalars belong to a triplet of the $SU(2)$ R-symmetry.
If one introduces another topological term associated with another $\vartheta_b$, this requires a different $SU(2)$ triplet which conflicts with the original one.
However,
since we regard the GLSM as string worldsheet theory and its UV completion, we believe that the broken supersymmetry must be restored in the IR limit.

\section{Conclusion and discussions} \label{sect:conclusion}
In this paper we have studied string winding corrections to the
five-brane, F-string and wave geometries.
We have explored classical DFT solutions of five-branes.
The solutions are classified according to the codimensions of the branes
in supergravity.
For example, applying the formal $O(d,d)$ transformations to the
NS5-brane of codimension four, we have found new solutions that depend
on multiple winding coordinates. These include the KK-monopole with one
winding, the $5^2_2$-brane with two windings and the $R$-brane with three
winding coordinates dependence.
The solution generating technique is also available
even for lower codimensional cases.
Starting from the H-monopole (codimension three) and smeared H-monopoles
(codimension two and one), we have shown various five-brane
solutions that involve winding coordinate dependency.
We have also written down codimension zero, space-filling five-branes
that are fully localized in the winding space.
Although they inevitably have trivial geometry (flat space-time)
in the supergravity language, they inherit non-trivial structures in the
winding space.

We have also worked out the winding solutions based on the DFT wave
solution. The analysis presented in this paper reveals that there are winding
corrections even for the F-string and wave solutions.
Although some of the solutions have been mentioned and found in the
literature \cite{Berkeley:2014nza, Berman:2014jsa, Bakhmatov:2016kfn,
Blair:2016xnn, Lust:2017jox},
our systematic analysis makes it clear that
for a given supergravity solution in the NSNS sector,
we can explicitly written down all the T-dualized solutions that depend on the winding coordinates.

In the latter half of this paper, we have discussed a physical
interpretation of the winding solutions.
We have separated out the winding sector from the ordinary supergravity
solutions. For solutions of less than two codimensions, the zero-winding
sectors diverge but they can be subtracted.
Except for the space-filling branes, we have elucidated the exponential
behaviors of the winding coordinate dependence of the solutions.
These characteristic structures suggest that they originate
from instanton effects.

We have then discussed an interpretation of the winding corrections as string
worldsheet instanton effects.
Based on the notion of the disk instantons in the single-centered
Taub-NUT space \cite{Okuyama:2005gx}, we have generalized it to the
$5^2_2$-brane geometry.
In an appropriate limit of parameters, a disk in the target space is
closed and the disk instantons are promoted to the well-defined
worldsheet instantons.
The map from the worldsheet to the target geometry is classified
according to the set of homotopy groups and the map is labeled by
integers associated with winding numbers.
Assuming that the Euclidean worldsheet action is minimized, we have
reproduced the correct exponential behavior that appeared in the DFT
winding solutions.
We have also given a discussion on the instanton effects based on the
GLSM language.
It is known that the worldsheet instantons are well described by the
GLSM formalism \cite{Tong:2002rq}.
The worldsheet instantons are represented by the ANO vortices in the GLSM.
Indeed, the winding corrections to the KK-monopole geometry
\eqref{eq:mod_codim3} is precisely reproduced by the path integral over
the vortex moduli space \cite{Harvey:2005ab}.
Related to this issue, two of the authors studied the instanton
corrections to the geometry of the exotic $5^2_2$-brane
\cite{Kimura:2013fda, Kimura:2013zva}.
Again it has been discussed that the ANO vortices induces
instanton corrections whose effect sums up to the one given in
\eqref{eq:codim2_reg}.
With these observations, it is plausible that the multiple winding dependence of the geometries
is captured by the ANO vortices associated with the $U(1)^N$ gauge symmetry in the GLSM.
We have a qualitative discussion on this issue.
Assuming the copies of the Abelian-Higgs model as a truncated model in a
limit of geometric parameters (this guarantees the instanton
interpretation), we have again reproduced the characteristic exponential
structure appearing in the harmonic function.

DFT is a framework where the $O(d,d)$ T-duality symmetry is manifest.
It is naturally generalized to the exceptional field theory (EFT)
\cite{Hohm:2013pua}, the U-duality covariant formulation, for which not
only the NSNS sector but also the RR sector of superstring theory are included.
Several classical solutions for branes including D-branes in EFT are
studied \cite{Berman:2014hna}.
Analogous to the DFT solutions presented here, it is discussed that there
are also EFT solutions that include ``dual coordinate dependence''
\cite{Bakhmatov:2017les}.
Related works on the winding corrections in the semi-flat elliptic metric is analyzed in \cite{Lust:2017jox, Achmed-Zade:2018rfc}.
Since the type II supergravity is manifestly included in
EFT, our analysis leads to the following conjecture:
\begin{quote}
If there is a classical solution to conventional type II supergravity theories,
one obtains associated DFT (EFT) solutions that depend on the
dual coordinates by formally applying T(U)-duality transformations to
 the solution.
Namely, there are always string winding corrections to every known
 supergravity solution.
\end{quote}
The appearance of the dual coordinates in EFT solutions is a consequence of
windings of various branes.
It is therefore natural to interpret it as instanton effects originating from wrapped branes on some cycles.

Since it is necessary to use the GLSM technique for the quantitative
analysis of the instanton effects, it is interesting to construct the
complete GLSM action that have $U(1)^N$ gauge symmetry.
It is also interesting to study the winding corrections to geometries
from the viewpoint of the worldvolume theories of exotic branes
\cite{Chatzistavrakidis:2013jqa,Kimura:2014upa,Kimura:2016anf, Blair:2017hhy}.
It is interesting to study the winding corrections to the
five-branes in heterotic theories \cite{Sasaki:2016hpp,Sasaki:2017yrs}.
We will come back to these issues in future works.

\subsection*{Acknowledgments}
The work of T.~K. is supported by the Iwanami-Fujukai Foundation.
The work of S.~S. is supported by the Japan Society for the Promotion of
Science (JSPS) KAKENHI Grant Number JP17K14294 and Kitasato University
Research Grant for Young Researchers.


\end{document}